\documentclass[10pt,letterpaper]{article}
\usepackage[margin=1.25in,dvips]{geometry}

\usepackage{amsthm}
\usepackage{times}

\usepackage{graphicx}
\usepackage{amssymb}
\usepackage{amsmath}
\usepackage{float}
\usepackage{url}
\usepackage{enumerate}

\usepackage[boxed,section]{algorithm}
\usepackage{algpseudocode}
\usepackage{color}

\newtheorem{theorem}{Theorem}[section]
\newtheorem{lemma}[theorem]{Lemma}
\newtheorem{corollary}[theorem]{Corollary}
\newtheorem{definition}[theorem]{Definition}
\newtheorem{claim}[theorem]{Claim}

\newcommand{\sabs}[1]{|#1|}
\newcommand{\abs}[1]{\left|#1\right|}
\newcommand{\floor}[1]{\left\lfloor#1\right\rfloor}
\newcommand{\ceil}[1]{\left\lceil#1\right\rceil}
\newcommand{\norm}[2]{\left \lVert#2\right \rVert_{#1}}
\newcommand{\snorm}[2]{\lVert#2\rVert_{#1}}

\newcommand{\wh}{\widehat}

{\makeatletter
 \gdef\xxxmark{%
   \expandafter\ifx\csname @mpargs\endcsname\relax 
     \expandafter\ifx\csname @captype\endcsname\relax 
       \marginpar{xxx}
     \else
       xxx 
     \fi
   \else
     xxx 
   \fi}
 \gdef\xxx{\@ifnextchar[\xxx@lab\xxx@nolab}
 \long\gdef\xxx@lab[#1]#2{{\bf [\xxxmark #2 ---{\sc #1}]}}
 \long\gdef\xxx@nolab#1{{\bf [\xxxmark #1]}}
}

\DeclareMathOperator*{\argmax}{arg\,max}
\DeclareMathOperator*{\argmin}{arg\,min}
\DeclareMathOperator*{\median}{median}
\DeclareMathOperator{\supp}{supp}
\DeclareMathOperator{\cdf}{cdf}
\DeclareMathOperator{\E}{\mathbb{E}}
\newcommand\R{\mathbb{R}}
\newcommand{\C}{{\mathbb C}}
\newcommand{\Z}{{\mathbb Z}}
\newcommand\eps{\epsilon}
\DeclareMathOperator{\err}{Err}
\DeclareMathOperator{\Err}{Err}

\newenvironment{Itemize}%
{\begin{itemize}%
\setlength{\itemsep}{0pt}%
\setlength{\topsep}{0pt}%
\setlength{\partopsep}{0pt}%
\setlength{\parskip}{0pt}}%
{\end{itemize}}
\newenvironment{Enumerate}%
{\begin{enumerate}%
\setlength{\itemsep}{0pt}%
\setlength{\topsep}{0pt}%
\setlength{\partopsep}{0pt}%
\setlength{\parskip}{0pt}}%
{\end{enumerate}}
\begin{document}

\title{Nearly Optimal Sparse Fourier Transform}

\author{Haitham Hassanieh \\MIT  \and  Piotr Indyk \\MIT  \and Dina Katabi \\MIT \and Eric Price\\MIT}
\date{{\tt \{haithamh,indyk,dk,ecprice\}@mit.edu}}


\maketitle
\begin{abstract}
  We consider the problem of computing the $k$-sparse approximation to
  the discrete Fourier transform of an $n$-dimensional signal. We
  show:
  \begin{itemize}
  \item An $O(k \log n)$-time randomized algorithm for the case where
    the input signal has at most $k$ non-zero Fourier coefficients,
    and
  \item An $O(k \log n \log(n/k))$-time randomized algorithm for
    general input signals.
  \end{itemize}
  Both algorithms achieve $o(n \log n)$ time, and thus improve over
  the Fast Fourier Transform, for any $k=o(n)$. They are the first
  known algorithms that satisfy this property. Also, if one assumes
  that the Fast Fourier Transform is optimal, the algorithm for the
  exactly $k$-sparse case is optimal for any $k = n^{\Omega(1)}$.

  We complement our algorithmic results by showing that any algorithm
  for computing the sparse Fourier transform of a general signal must
  use at least $\Omega(k \log (n/k) / \log \log n)$ signal samples,
  even if it is allowed to perform \emph{adaptive} sampling.
\end{abstract}


\section{Introduction}\label{sec:intro}
The discrete Fourier transform (DFT) is one of the most important and
widely used computational tasks. Its applications are broad and
include signal processing, communications, and audio/image/video
compression.  Hence, fast algorithms for DFT are highly
valuable. Currently, the fastest such algorithm is the Fast Fourier
Transform (FFT), which computes the DFT of an $n$-dimensional signal
in $O(n \log n)$ time. The existence of DFT algorithms faster
  than FFT is one of the central questions in the theory of
  algorithms.

A general algorithm for computing the
exact DFT must take time at least proportional to its output size,
i.e., $\Omega(n)$.
In many applications, however, most of the Fourier coefficients of a
signal are small or equal to zero, i.e., the output of the DFT is
(approximately) {\em sparse}. This is the case for video signals,
where a typical 8x8 block in a video frame has on average 7
non-negligible frequency coefficients (i.e., 89\% of the coefficients
are negligible)~\cite{Anantha}.  Images and audio data are equally
sparse. This sparsity provides the rationale underlying compression
schemes such as MPEG and JPEG.  Other sparse signals appear in
computational learning theory~\cite{KM,LMN}, analysis of Boolean
functions~\cite{KKL,ODon}, compressed
sensing~\cite{Don,CRT}, multi-scale analysis~\cite{DRZ}, similarity search in databases~\cite{AFS},
spectrum sensing for wideband channels~\cite{app-2}, and datacenter
monitoring~\cite{app-1}. 

For sparse signals, the $\Omega(n)$ lower bound for the complexity of
DFT no longer applies.  If a signal has a small number $k$ of non-zero
Fourier coefficients -- the {\em exactly $k$-sparse} case -- the output
of the Fourier transform can be represented succinctly using only $k$
coefficients. Hence, for such signals, one may hope for a DFT
algorithm whose runtime is sublinear in the signal size, $n$.  Even
for a general $n$-dimensional signal $x$ -- the {\em general case} -- one
can find an algorithm that computes the best {\em k-sparse
  approximation} of its Fourier transform, $\wh{x}$, in sublinear
time.  The goal of such an algorithm is to compute an approximation
vector $\wh{x}'$ that satisfies the following {\em $\ell_2/\ell_2$
  guarantee}:
\begin{equation}
\label{e:l2l2}
\| \wh{x}-\wh{x}'\|_2 \le C \min_{k \text{-sparse } y }  \|\wh{x}-y\|_2,
\end{equation}
where $C$ is some approximation factor and the minimization is over
$k$-sparse signals.  We allow the algorithm to be \emph{randomized},
and only succeed with constant (say, 2/3) probability.

%

The past two decades have witnessed significant advances in sublinear
sparse Fourier algorithms.  The first such algorithm (for the Hadamard
transform) appeared in~\cite{KM} (building on~\cite{GL}). Since then,
several sublinear sparse Fourier algorithms for complex inputs have been
discovered \cite{Man,GGIMS,AGS,GMS,Iw,Ak,HIKP}. These algorithms
provide\footnote{The algorithm of~\cite{Man}, as stated in the paper,
  addresses only the exactly $k$-sparse case.  However, it can be
  extended to the general case using relatively standard techniques.}
the guarantee in Equation~\eqref{e:l2l2}.\footnote{All of the above
  algorithms, as well as the algorithms in this paper, need to make
  some assumption about the precision of the input; otherwise, the
  right-hand-side of the expression in Equation~\eqref{e:l2l2}
  contains an additional additive term.  See Preliminaries for more
  details.}

The main value of these algorithms is that they outperform FFT's
runtime for sparse signals. For very sparse signals, the fastest
algorithm is due to~\cite{GMS} and has $O(k \log^c(n) \log(n/k))$ runtime, for some\footnote{The paper does not estimate the exact value of $c$. We estimate  that $c \approx 3$.} 
$c>2$.  This algorithm outperforms FFT for any $k$ smaller than
$\Theta(n/\log^a n)$ for some $a>1$. For less sparse signals, the
fastest algorithm is due to~\cite{HIKP}, and has $O(\sqrt{nk}
\log^{3/2} n)$ runtime. This algorithm outperforms FFT for any $k$
smaller than $\Theta(n/\log n)$.

Despite impressive progress on sparse DFT, the state of the art
suffers from two main limitations:
 \begin{Enumerate}
\item None of the existing algorithms improves over FFT's runtime for
  the whole range of sparse signals, i.e., $k=o(n)$.

 \item Most of the aforementioned algorithms are quite complex, and
   suffer from large ``big-Oh'' constants (the algorithm of \cite{HIKP}
   is an exception, but has a running time that is polynomial in
   $n$). 
  \end{Enumerate}

\paragraph{Results.} 
In this paper, we address these limitations by presenting two new
algorithms for the sparse Fourier transform. We require that the length
$n$ of the input signal is a power of 2.
We show:
   \begin{Itemize}
   \item An $O(k \log n)$-time algorithm for the exactly $k$-sparse
     case, and
   \item An $O(k \log n \log(n/k))$-time algorithm for the general case.   
    \end{Itemize}
 The key property of both algorithms is their ability to achieve $o(n
 \log n)$ time, and thus improve over the FFT, for {\em any}
 $k=o(n)$. 
These algorithms are the first known algorithms that satisfy this property.
 Moreover, if one assume that FFT is optimal and hence the DFT cannot be
 computed in less than $O(n \log n)$ time, the algorithm for the exactly
 $k$-sparse case is {\em optimal}\footnote{One also needs to assume
   that $k$ divides $n$. See Section~\ref{s:red} for more details.} as
 long as $k=n^{\Omega(1)}$. Under the
 same assumption, the result for the general case is at most one $\log
 \log n$ factor away from the optimal runtime for the case of
 ``large'' sparsity $k=n/\log^{O(1)} n$.

 Furthermore, our algorithm for the exactly sparse case (depicted as
 Algorithm~\ref{a:exact} on page 5) is quite simple and has low big-Oh
 constants.  In particular, our preliminary implementation of a
 variant of this algorithm is faster than FFTW, a highly efficient
 implementation of the FFT, for $n=2^{22}$ and $k\le 2^{17}$~\cite{sFFT}. In
 contrast, for the same signal size, prior algorithms were faster than
 FFTW only for $k \le 2000$~\cite{HIKP}.\footnote{Note that both numbers
   ($k\le 2^{17}$ and $k \le 2000$) are for the exactly k-sparse case. The algorithm
   in~\cite{HIKP} can deal with the general case, but the
   empirical runtimes are higher.}


  
 We complement our algorithmic results by showing that any algorithm
 that works for the general case must use at least $\Omega(k \log
 (n/k)/\log \log n)$ samples from $x$.  The lower bound uses
 techniques from~\cite{PW}, which shows a lower bound of $\Omega(k
 \log (n/k))$ for the number of {\em arbitrary} linear measurements
 needed to compute the $k$-sparse approximation of an $n$-dimensional
 vector $\wh{x}$.  In comparison to~\cite{PW}, our bound is slightly
 worse but it holds even for {\em adaptive} sampling, where the
 algorithm selects the samples based on the values of the previously
 sampled coordinates.\footnote{Note that if we allow {\em arbitrary}
   adaptive linear measurements of a vector $\wh{x}$, then its
   $k$-sparse approximation can be computed using only $O(k \log \log
   (n/k))$ samples~\cite{IPW}. Therefore, our lower bound holds only
   where the measurements, although adaptive, are limited to those
   induced by the Fourier matrix. This is the case when we want to
   compute a sparse approximation to $\wh{x}$ from samples of $x$.}
 Note that our algorithms are {\em non-adaptive}, and thus limited by
 the more stringent lower bound of~\cite{PW}.

   
  \paragraph{Techniques -- overview.} 
We start with an overview of the techniques used in prior works.  At a
high level, sparse Fourier algorithms work by binning the Fourier
coefficients into a small number of bins.  Since the signal is sparse
in the frequency domain, each bin is likely\footnote{One can randomize
  the positions of the frequencies by sampling the signal in time
  domain appropriately~\cite{GGIMS,GMS}. See Preliminaries for the
  description.} to have only one large coefficient, which can then be
located (to find its position) and estimated (to find its value). The
binning has to be done in sublinear time, and thus these algorithms
bin the Fourier coefficients using an $n$-dimensional filter vector
$G$ that is concentrated both in time and frequency. That is, $G$ is
zero except at a small {\em number} of time coordinates, and its
Fourier transform $\hat{G}$ is negligible except at a small {\em
  fraction} (about $1/k$) of the frequency coordinates, representing
the filter's ``pass'' region.  Each bin essentially receives only the
frequencies in a narrow range corresponding to the pass region of the
(shifted) filter $\hat{G}$, and the pass regions corresponding to
different bins are disjoint.  In this paper, we use filters introduced
in~\cite{HIKP}.  Those filters (defined in more detail in
Preliminaries) have the property that the value of $\hat{G}$ is
``large'' over a constant fraction of the pass region, referred to as
the ``super-pass'' region.  We say that a coefficient is ``isolated''
if it falls into a filter's super-pass region and no other coefficient
falls into filter's pass region.  Since the super-pass region of our
filters is a constant fraction of the pass region, the probability of
isolating a coefficient is constant.  

To achieve the stated running times, we need a fast method
for locating and estimating isolated coefficients. Further, our
algorithm is iterative, so we also need a fast method for updating the
signal so that identified coefficients are not considered in future
iterations. Below, we describe these methods in more
detail.

\paragraph{New techniques -- location and estimation.} 
Our location and estimation methods depends on whether we handle the
exactly sparse case or the general case. In the exactly sparse case,
we show how to estimate the position of an isolated Fourier
coefficient using only two samples of the filtered
signal. Specifically, we show that the phase difference between the
two samples is linear in the index of the coefficient, and hence we
can recover the index by estimating the phases. This approach is
inspired by the frequency offset estimation in orthogonal frequency
division multiplexing (OFDM), which is the modulation method used in
modern wireless technologies (see~\cite{HT}, Chapter 2).
    
In order to design an algorithm\footnote{We note that although the
  two-sample approach employed in our algorithm works in theory only
  for the exactly $k$-sparse case, our preliminary experiments show
  that using a few more samples to estimate the phase works
  surprisingly well even for general signals.}  for the general case,
we employ a different approach. Specifically, we can use two samples
to estimate (with constant probability) individual bits of the index
of an isolated coefficient. Similar approaches have been employed in
prior work.
 However, in those papers, the index was
recovered bit by bit, and one needed $\Omega(\log \log n)$ samples per
bit to recover {\em all} bits correctly with constant probability.  In
contrast, in this paper we recover the index one {\em block of bits}
at a time, where each block consists of $O(\log \log n)$ bits. This
approach is inspired by the fast sparse recovery algorithm
of~\cite{GLPS}. Applying this idea in our context, however, requires
new techniques. The reason is that, unlike in~\cite{GLPS}, we do not
have the freedom of using arbitrary ``linear measurements'' of the
vector $\hat{x}$, and we can only use the measurements induced by the
Fourier transform.\footnote{In particular, the method of~\cite{GLPS}
  uses measurements corresponding to a random error correcting code.}
 As a result, the extension from ``bit recovery'' to
``block recovery'' is the most technically involved part of the
algorithm. Section~\ref{ss:int} contains further intuition on this part.

\paragraph{New techniques -- updating the signal.} 
The aforementioned techniques recover the position and the value of
any isolated coefficient. However, during each filtering step, each
coefficient becomes isolated only with constant probability. Therefore,
the filtering process needs to be repeated to ensure that each
coefficient is correctly identified. In~\cite{HIKP}, the algorithm
simply performs the filtering $O(\log n)$ times and uses the median
estimator to identify each coefficient with high probability. This,
however, would lead to a running time of $O(k \log^2 n)$ in the
$k$-sparse case, since each filtering step takes $k \log n$ time.

One could reduce the filtering time by subtracting the identified
coefficients from the signal. In this way, the number of non-zero
coefficients would be reduced by a constant factor after each
iteration, so the cost of the first iteration would dominate the total
running time.  Unfortunately, subtracting the recovered coefficients
from the signal is a computationally costly operation, corresponding
to a so-called {\em non-uniform} DFT (see~\cite{GST} for details). Its
cost would override any potential savings.

In this paper, we introduce a different approach: instead of
subtracting the identified coefficients from the {\em signal}, we
subtract them directly from the {\em bins} obtained by filtering the
signal. The latter operation can be done in time linear in the number
of subtracted coefficients, since each of them ``falls'' into only one
bin. Hence, the computational costs of each iteration can be
decomposed into two terms, corresponding to filtering the original
signal and subtracting the coefficients. For the exactly sparse case
these terms are as follows:
\begin{Itemize}
\item The cost of filtering the original signal is 
  $O(B \log n)$, where $B$ is the number of bins. $B$ is set to
  $O(k')$, where $k'$ is the the number of yet-unidentified
  coefficients. Thus, initially $B$ is equal to $O(k)$, but its value
  decreases by a constant factor after each iteration.
\item The cost of subtracting the identified coefficients from the
  bins  is  $O(k)$. 
\end{Itemize}
Since the number of iterations is $O(\log k)$, and the cost of
filtering is dominated by the first iteration, the total
running time is $O(k \log n)$ for the exactly sparse case.

For the general case, we need to set $k'$ and $B$ more carefully to
obtain the desired running time.  The cost of each iterative step is
multiplied by the number of filtering steps needed to compute the
location of the coefficients, which is $\Theta(\log (n/B))$.  If we
set $B = \Theta(k')$, this would be $\Theta(\log n)$ in most
iterations, giving a $\Theta(k \log^2 n)$ running time.  This is too
slow when $k$ is close to $n$.  We avoid this by decreasing $B$ more
slowly and $k'$ more quickly.  In the $r$-th iteration, we set $B = k
/ \text{poly}(r)$.  This allows the total number of bins to remain
$O(k)$ while keeping $\log (n/B)$ small---at most $O(\log \log k)$
more than $\log(n/k)$.  Then, by having $k'$ decrease according to $k'
= k/r^{\Theta(r)}$ rather than $k / 2^{\Theta(r)}$, we decrease the
number of rounds to $O(\log k / \log \log k)$.  Some careful analysis
shows that this counteracts the $\log \log k$ loss in the $\log(n/B)$
term, achieving the desired $O(k \log n \log (n/k))$ running time.



   

\paragraph{Organization of the paper.}  In Section~\ref{sec:defs}, we
give notation and definitions used throughout the paper.
Sections~\ref{sec:inf} and~\ref{sec:general} give our algorithm in the
exactly $k$-sparse and the general case, respectively.
Section~\ref{s:red} gives the reduction to the exactly $k$-sparse case
from a $k$-dimensional DFT.  Section~\ref{sec:lower} gives the sample
complexity lower bound for the general case.
Section~\ref{sec:efficientwindow} describes how to efficiently
implement our filters.  Finally, Section~\ref{sec:open} discusses open
problems arising from this work.

\newcommand{\hs}{h_{\sigma,b}}
\newcommand{\hsr}{h_{\sigma_r,b_r}}
\newcommand{\ps}{\pi_{\sigma,b}}
\newcommand{\os}{o_{\sigma,b}}

\section{Preliminaries}\label{sec:defs}
This section introduces the notation, assumptions, and definitions
used in the rest of this paper.


\paragraph{Notation.} We use $[n]$ to denote the set $\{ 1, \dotsc,
n\}$, and define $\omega = e^{-2\pi \mathbf{i} / n}$ to be an $n$th
root of unity.  For any complex number $a$, we use $\phi(a) \in [0,
2\pi]$ to denote the {\em phase} of $a$.  For a complex number $a$ and
a real positive number $b$, the expression $a \pm b$ denotes a complex
number $a'$ such that $\abs{a-a'} \le b$.  For a vector $x \in \C^n$,
its support is denoted by $\supp(x) \subset [n]$.  We use
$\norm{0}{x}$ to denote $\abs{\supp(x)}$, the number of non-zero
coordinates of $x$.  Its Fourier spectrum is denoted by $\wh{x}$, with
\[
\wh{x}_i = \frac{1}{\sqrt{n}}\sum_{j\in [n]} \omega^{i j} x_j.
\]
For a vector of length $n$, indices should be interpreted modulo $n$,
so $x_{-i} = x_{n-i}$.  This allows us to define \emph{convolution}
\[
(x * y)_i = \sum_{j \in [n]} x_{j} y_{i-j}
\]
and the \emph{coordinate-wise product} $(x \cdot y)_i = x_i y_i$, so
$\wh{x \cdot y} = \wh{x} * \wh{y}$.

When $i \in \Z$ is an index into an $n$-dimensional vector, sometimes
we use $\abs{i}$ to denote $\min_{j \equiv i \pmod{n}} \abs{j}$.

\paragraph{Definitions.} 
The paper uses two tools introduced in previous papers: (pseudorandom) spectrum permutation~\cite{GGIMS,GMS,GST} and flat filtering windows~\cite{HIKP}.

\begin{definition}
  Suppose $\sigma^{-1}$ exists mod $n$.  We define the
  \emph{permutation} $P_{\sigma, a, b}$ by
  \[
  (P_{\sigma,a,b}x)_i = x_{\sigma (i - a)} \omega^{\sigma bi}.
  \]
  We also define $\ps(i) = \sigma(i-b) \bmod n$.
\end{definition}
\begin{claim}\label{claim:perm}
  $\wh{P_{\sigma,a,b}x}_{\ps(i)} = \wh{x}_i\omega^{a\sigma i}$.
\end{claim}
\begin{proof}
  \begin{align*}
    \wh{P_{\sigma,a,b}x}_{\sigma(i - b)} 
    &= \frac{1}{\sqrt{n}} \sum_{j \in [n]} \omega^{\sigma(i-b)j}(P_{\sigma,a,b}x)_j\\
    &= \frac{1}{\sqrt{n}} \sum_{j \in [n]} \omega^{\sigma(i-b)j}x_{\sigma (j-a)}\omega^{\sigma bj}\\
    &= \omega^{a\sigma i}\frac{1}{\sqrt{n}} \sum_{j \in [n]} \omega^{i \sigma (j-a)}x_{\sigma (j-a)}\\
    &= \wh{x}_i\omega^{a\sigma i}.
  \end{align*}
\end{proof}

\begin{definition}
  We say that $(G, \wh{G'}) = (G_{B,\delta,\alpha},
  \wh{G'}_{B,\delta,\alpha}) \in \R^n \times \R^n$ is a \emph{flat window
    function} with parameters $B \geq 1$, $\delta > 0$, and $\alpha >
  0$ if $\abs{\supp(G)} = O(\frac{B}{\alpha}\log(n/\delta))$ and
  $\wh{G'}$ satisfies
  \begin{itemize}
  \item $\wh{G'}_i = 1$ for $\abs{i} \leq (1-\alpha)n/(2B)$
  \item $\wh{G'}_i = 0$ for $\abs{i} \geq n/(2B)$
  \item $\wh{G'}_i \in [0, 1]$ for all $i$
  \item $\norm{\infty}{\wh{G'}-\wh{G}} < \delta$.
  \end{itemize}
\end{definition}

The above notion corresponds to the $(1/(2B), (1-\alpha)/(2B), \delta,
O(B/\alpha \log (n/\delta))$-flat window function in~\cite{HIKP}. In
Section~\ref{sec:efficientwindow} we give efficient constructions of
such window functions, where $G$ can be computed in
$O(\frac{B}{\alpha} \log (n/\delta))$ time and for each $i$,
$\wh{G'}_i$ can be computed in $O(\log(n/\delta))$ time. Of course,
for $i \notin [(1-\alpha)n/(2B), n/(2B)]$, $\wh{G'}_i \in \{0,1\}$ can
be computed in $O(1)$ time.

The fact that $\wh{G'}_i$ takes $\omega(1)$ time to compute for $i \in
[(1-\alpha)n/(2B), n/(2B)]$ will add some complexity to our algorithm
and analysis.  We will need to ensure that we rarely need to compute
such values.  A practical implementation might find it more convenient
to precompute the window functions in a preprocessing stage, rather
than compute them on the fly.

We use the following lemma from~\cite{HIKP}:
\begin{lemma}[Lemma 3.6 of~\cite{HIKP}]\label{lemma:limitedindependence}
  If $j \neq 0$, $n$ is a power of two, and $\sigma$ is a uniformly
  random odd number in $[n]$, then $\Pr[\sigma j \in [-C, C] \pmod n ]
  \leq 4C/n$.
\end{lemma}

\paragraph{Assumptions.} Through the paper, we require that $n$, the
dimension of all vectors, is an integer power of $2$.  We also make
the following assumptions about the precision of the vectors $\wh{x}$:
\begin{Itemize}
\item For the exactly $k$-sparse case, we assume that $\wh{x}_i \in
  \{-L, \ldots, L\}$ for some precision parameter $L$. To simplify the
  bounds, we assume that $L =n^{O(1)}$; otherwise the $\log n$ term in
  the running time bound is replaced by $\log L$.
\item For the general case, we only achieve Equation~\eqref{e:l2l2} if
  $\norm{2}{\wh{x}} \leq n^{O(1)} \cdot \min_{k \text{-sparse } y }
  \norm{2}{\wh{x}-y}$. In general, for any parameter $\delta > 0$ we
  can add $\delta \norm{2}{\wh{x}}$ to the right hand side of
  Equation~\eqref{e:l2l2} and run in time $O(k \log (n/k) \log (n/\delta))$.
\end{Itemize}

\section{Algorithm for the exactly sparse case}\label{sec:inf}

In this section we assume $\wh{x}_i \in \{-L, \dotsc, L\}$, where $L \le
n^c$ for some constant $c>0$, and $\wh{x}$ is $k$-sparse.  We choose
$\delta = 1/(4 n^2 L)$.  The algorithm (\textsc{NoiselessSparseFFT})
is described as Algorithm~\ref{a:exact}.  The algorithm has three
functions:
\begin{itemize}
\item \textsc{HashToBins}.  This permutes the spectrum of $\wh{x-z}$
  with $P_{\sigma, a, b}$, then ``hashes'' to $B$ bins.  The guarantee
  will be described in Lemma~\ref{lemma:hashtobins}.
\item \textsc{NoiselessSparseFFTInner}.  Given time-domain access to
  $x$ and a sparse vector $\wh{z}$ such that $\wh{x-z}$ is
  $k'$-sparse, this function finds ``most'' of $\wh{x-z}$.
\item \textsc{NoiselessSparseFFT}.  This iterates
  \textsc{NoiselessSparseFFTInner} until it finds $\wh{x}$ exactly.
\end{itemize}


\begin{algorithm}[t!]
 \caption{Exact $k$-sparse recovery}\label{a:exact}
  \begin{algorithmic}
    \Procedure{HashToBins}{$x$, $\wh{z}$, $P_{\sigma, a, b}$, $B$, $\delta$, $\alpha$}
    \State Compute $\wh{y}_{jn/B}$ for $j \in [B]$, where $y = G_{B,\alpha,\delta} \cdot (P_{\sigma,a,b}x)$ 
    \State Compute $\wh{y'}_{jn/B} = \wh{y}_{jn/B} -  (\wh{G'_{B,\alpha,\delta}} * \wh{P_{\sigma,a,b}z} )_{jn/B} $ for $j \in [B]$ 
    \State \Return $\wh{u}$ given by $\wh{u}_j = \wh{y'}_{jn/B}$.
    \EndProcedure

    \Procedure{NoiselessSparseFFTInner}{$x$, $k'$, $\wh{z}$, $\alpha$}

    \State Let $B$ = $k'/\beta$, for sufficiently small constant $\beta$.
    \State Let $\delta = 1/(4n^2L)$.
    \State Choose  $\sigma$ uniformly at random from the set of odd numbers in $[n]$.
    \State Choose $b$ uniformly at random from $[n]$.
       \State $\wh{u} \gets \textsc{HashToBins}(x, \wh{z}, P_{\sigma,0,b}, B, \delta, \alpha)$.
    \State $\wh{u}' \gets \textsc{HashToBins}(x, \wh{z}, P_{\sigma,1,b}, B, \delta, \alpha)$.
     \State $\wh{w} \gets 0$. 
    \State Compute $J=\{j: |\wh{u}_j| > 1/2\}$.
     \For{$j \in J$}
    \State $a \gets \wh{u}_{j} / \wh{u}'_{j}$.
   \State $i  \gets \sigma^{-1}( \text{round}(\phi(a)\frac{n}{2\pi})) \bmod n$. \Comment{$\phi(a)$ denotes the phase of $a$.}
    \State $v \gets \text{round}(\wh{u}_{j})$.
   \State $\wh{w}_i \gets v$. 
    \EndFor 
    \State \Return $\wh{w}$
    \EndProcedure
    \Procedure{NoiselessSparseFFT}{$x$, $k$} 
    \State $\wh{z} \gets 0$
    \For{$t \in 0,1, \dotsc, \log k$}
    \State $k_t \gets k/2^t$, $\alpha_t \gets \Theta(2^{-t})$.
    \State $\wh{z} \gets \wh{z} + \textsc{NoiselessSparseFFTInner}(x, k_t, \wh{z}, \alpha_t)$. 

    \EndFor
    \State \Return $\wh{z}$
    \EndProcedure
  \end{algorithmic}
\end{algorithm}

We analyze the algorithm ``bottom-up'', starting from the lower-level procedures.

\paragraph{Analysis of \textsc{NoiselessSparseFFTInner} and \textsc{HashToBins}.}

\newcommand{\Ec}{E_{coll}}
\newcommand{\Eo}{E_{off}}
\newcommand{\En}{E_{noise}}

For any execution of \textsc{NoiselessSparseFFTInner}, define the support
$S=\supp( \wh{x} -\wh{z} )$. Recall that $\ps(i) = \sigma (i-b) \bmod
n$.  Define $\hs(i) = \text{round}( \ps(i) B/n)$ and $\os(i)=\ps(i) -
\hs(i) n/B$. Note that therefore $\abs{\os(i)} \leq n/(2B)$.  We will
refer to $\hs(i)$ as the ``bin'' that the frequency $i$ is mapped
into, and $\os(i)$ as the ``offset''.  For any $i \in S$ define two
types of events associated with $i$ and $S$ and defined over the
probability space induced by $\sigma$ and $b$:
\begin{Itemize}
\item ``Collision'' event $\Ec(i)$: holds iff $\hs(i) \in \hs(S\setminus\{i\})$, and
\item ``Large offset'' event $\Eo(i)$: holds iff $|\os(i)| \ge (1-\alpha) n/(2B)$.
\end{Itemize}

\begin{claim} 
\label{c:coll}
For any $i \in S$, the event $\Ec(i)$ holds with probability at most $4|S|/B$.
\end{claim}
\begin{proof}
  Consider distinct $i, j \in S$.  By
  Lemma~\ref{lemma:limitedindependence},
  \begin{align*}
    \Pr[\hs(i) = \hs(j)] 
    &\leq
    \Pr[\ps(i) - \ps(j) \bmod n \in [-n/B, n/B]]\\
    &= \Pr[\sigma(i-j)
    \bmod n \in [-n/B,n/B]] \\&\leq 4/B.
  \end{align*}
  By a union bound over $j \in S$, $\Pr[\Ec(i)] \leq 4\abs{S}/B$.
\end{proof}

\begin{claim}
\label{c:off}
For any $i \in S$, the event $\Eo(i)$ holds with probability at most $\alpha$.
\end{claim}
\begin{proof}
  Note that $\os(i) \equiv \ps(i) \equiv \sigma(i - b) \pmod {n/B}$.
  For any odd $\sigma$ and any $l \in [n/B]$, we have that $\Pr_b
  [\sigma(i-b) \equiv l \pmod{n/B}] = B/n$.  Since only $\alpha n / B$
  offsets $\os(i)$ cause $\Eo(i)$, the claim follows.
\end{proof}

\begin{lemma}\label{lemma:hashtobins}
  Suppose $B$ divides $n$.  The output $\wh{u}$ of \textsc{HashToBins}
  satisfies
  \[
  \wh{u}_j = \sum_{\hs(i) = j} \wh{(x-z)}_i \wh{(G'_{B,\delta,\alpha})}_{-\os(i)} \omega^{a\sigma i} \pm \delta \norm{1}{\wh{x}}.
  \]
  Let $\zeta = \abs{\{i \in \supp(\wh{z}) \mid \Eo(i)\}}$.  The
  running time of \textsc{HashToBins} is $O(\frac{B}{\alpha}\log
  (n/\delta) + \norm{0}{\wh{z}} + \zeta \log (n/\delta))$.
\end{lemma}
\begin{proof}
  Define the flat window functions $G = G_{B,\delta,\alpha}$ and
  $\wh{G'}=\wh{G'}_{B,\delta,\alpha}$.  We have
  \begin{align*}
    \wh{y} &= \wh{G \cdot P_{\sigma,a,b}}x = \wh{G} * \wh{P_{\sigma,a,b}x}\\
    \wh{y'} &= \wh{G'} * \wh{P_{\sigma,a,b}(x-z)} + (\wh{G}-\wh{G'}) * \wh{P_{\sigma,a,b}x}
  \end{align*}
  By Claim~\ref{claim:perm}, the coordinates of $\wh{P_{\sigma,a,b}x}$
  and $\wh{x}$ have the same magnitudes, just different ordering and
  phase.  Therefore
  \[
  \norm{\infty}{(\wh{G}-\wh{G'}) * \wh{P_{\sigma,a,b}x}} \leq \norm{\infty}{\wh{G}-\wh{G'}}\norm{1}{\wh{P_{\sigma,a,b}x}} \leq \delta\norm{1}{\wh{x}} 
  \]
  and hence
  \begin{align*}
    \wh{u}_j
    = \wh{y'}_{jn/B}
    &= \sum_{\abs{l} < n/(2B)} \wh{G'}_{-l} \wh{(P_{\sigma,a,b}(x-z))}_{jn/B+l}  \pm \delta\norm{1}{\wh{x}}\\
    &= \sum_{\abs{\ps(i)-jn/B} < n/(2B)} 
    \wh{G'}_{jn/B-\ps(i)} \wh{(P_{\sigma,a,b}(x-z))}_{\ps(i)}  \pm \delta\norm{1}{\wh{x}}\\
    &= \sum_{\hs(i) = j} \wh{G'}_{-\os(i)} \wh{(x-z)}_{i}\omega^{a\sigma i}  \pm \delta\norm{1}{\wh{x}}
  \end{align*}
  as desired.

  We can compute \textsc{HashToBins} via the following method:
  \begin{enumerate}
  \item Compute $y$ with $\norm{0}{y} = O(\frac{B}{\alpha} \log (n/\delta))$ in $O(\frac{B}{\alpha} \log (n/\delta))$ time.
  \item Compute $v \in \C^B$ given by $v_i = \sum_{j} y_{i+jB}$.
  \item Because $B$ divides $n$, by the definition of the Fourier
    transform (see also Claim~3.7 of~\cite{HIKP}) we have
    $\wh{y}_{jn/B} = \wh{v}_j$ for all $j$.  Hence we can compute it
    with a $B$-dimensional FFT in $O(B \log B)$ time.
  \item For each coordinate $i \in \supp(\wh{z})$, decrease
    $\wh{y}_{\frac{n}{B}\hs(i)}$ by
    $\wh{G'}_{-\os(i)}\wh{z}_i\omega^{a\sigma i}$.  This takes
    $O(\norm{0}{\wh{z}} + \zeta \log(n/\delta))$ time, since
    computing $\wh{G'}_{-\os(i)}$ takes $O(\log(n/\delta))$ time if
    $\Eo(i)$ holds and $O(1)$ otherwise.
  \end{enumerate}
\vspace{-.2in}
\end{proof}

\begin{lemma}
\label{l:est}
Consider any $i \in S$ such that neither $\Ec(i)$ nor $\Eo(i)$ holds. 
Let $j = \hs(i)$.
Then 
\[ \text{round}(\phi(\wh{u}_j/\wh{u}'_j)) \frac{n}{2\pi}) = \sigma i \pmod{n},\]
\[ \text{round}(\wh{u}_j )= \wh{x}_i - \wh{z}_i, \]
and $j \in J$.
\end{lemma}

\begin{proof}
  We know that $\norm{1}{\wh{x}} \leq k\norm{\infty}{\wh{x}} \leq kL < nL$.  Then by
  Lemma~\ref{lemma:hashtobins} and $\Ec(i)$ not holding,
  \[
  \wh{u}_j = \wh{(x-z)}_i\wh{G'}_{- \os(i)}  \pm \delta n L.
  \]
  Because $\Eo(i)$ does not hold, $\wh{G'}_{-\os(i)} = 1$, so
  \begin{align}\label{e:ubound}
    \wh{u}_j = \wh{(x-z)}_i \pm \delta n L.
  \end{align}
  Similarly,
  \[
  \wh{u}_j' = \wh{(x-z)}_i\omega^{\sigma i}  \pm \delta n L
  \]
  Then because $\delta n L < 1 \leq \abs{\wh{(x-z)}_i}$, the phase is
  \[
  \phi(\wh{u}_j) = 0 \pm \sin^{-1} (\delta n L) = 0 \pm 2\delta n L
  \]
  and $\phi(\wh{u}_j') = -\sigma i\frac{2\pi}{n} \pm 2\delta n L$.  Thus
  $\phi(\wh{u}_j/\wh{u}'_j) = \sigma i\frac{2\pi}{n} \pm 4\delta n L =
  \sigma i\frac{2\pi}{n} \pm 1/n$ by the choice of $\delta$.  Therefore
  \[
  \text{round}(\phi(\wh{u}_j/\wh{u}'_j ) \frac{n}{2\pi}) = \sigma i \pmod n.
  \]

  Also, by Equation~\eqref{e:ubound}, $\text{round}(\wh{u}_j ) =
  \wh{x}_i-\wh{z}_i$.  Finally, $\abs{\text{round}(\wh{u}_j)} =
  \abs{\wh{x}_i-\wh{z}_i} \geq 1$, so $|\wh{u}_j| \ge 1/2$.  Thus $j
  \in J$.
\end{proof}

For  each invocation of \textsc{NoiselessSparseFFTInner}, let $P$ be the the set of all pairs $ (i,v)$ for which the command $\wh{w}_i \gets v$ was executed. 
Claims~\ref{c:coll} and~\ref{c:off} and Lemma~\ref{l:est} together guarantee that for each $i \in S$ the probability that $P$ does not contain the pair  $(i, (\wh{x}-\wh{z})_i)$ is at most $4|S|/B + \alpha$. 
We complement this observation with the following claim.

\begin{claim}
\label{c:onlyS}
For any $j \in J$ we have $j \in \hs(S)$. Therefore, $|J|  = |P| \le |S|$.
\end{claim}
\begin{proof}
  Consider any $j \notin \hs(S)$.  From Equation~\eqref{e:ubound} in the proof of
  Lemma~\ref{l:est} it follows that $|\wh{u}_j| \le \delta n L < 1/2$.
\end{proof}

\begin{lemma}
\label{l:exp}
Consider an execution of \textsc{NoiselessSparseFFTInner}, and let 
$S=\supp(\wh{x}-\wh{z})$. 
If  $|S| \le k'$, then 
\[ E [\|\wh{x}-\wh{z} - \wh{w}\|_0] \le 8(\beta+\alpha)|S| .\]
\end{lemma}

\begin{proof}
Let $e$ denote the number of coordinates $i \in S$ for which either $\Ec(i)$ or $\Eo(i)$ holds. 
Each such coordinate might not appear in $P$ with the correct value, leading to an incorrect value of $\wh{w}_i$.
In fact, it might result in an arbitrary pair $(i',v')$ being added to $P$, which in turn could lead to an incorrect value of 
 $\wh{w}_{i'}$. By Claim~\ref{c:onlyS} these are the only ways that $\wh{w}$ can be assigned an incorrect value.
 Thus we have
 \[
 \|\wh{x}-\wh{z} - \wh{w}\|_0 \le 2e.
 \]
 Since $E[e] \le  (4|S|/B + \alpha)|S| \le (4\beta+\alpha)|S| $, the lemma follows.
\end{proof}

\paragraph{Analysis of \textsc{NoiselessSparseFFT}.}

Consider the $t$th iteration of the procedure, and define $S_t=\supp(\wh{x}-\wh{z})$ where $\wh{z}$ denotes the value of the variable at the beginning of loop.
Note that $|S_0| =|\supp(\wh{x})| \le k$. 

We also define an indicator variable $I_t$ which is equal to $0$ iff $|S_t|/|S_{t-1}| \le 1/8$. If $I_t=1$ we say the the $t$th iteration was not {\em successful}.
Let $\gamma = 8 \cdot 8 (\beta+\alpha)$.
 From Lemma~\ref{l:exp} it follows that $\Pr[I_t=1 \mid |S_{t-1}|\le k/2^{t-1}] \le \gamma$. From Claim~\ref{c:onlyS} it follows that even if the $t$th iteration is not successful, then 
$|S_{t}|/|S_{t-1}| \le 2$.  

For any $t \ge 1$, define an event $E(t)$ that occurs iff $\sum_{i=1}^t I_i \ge t/2$. 
Observe that if none of the events $E(1) \ldots E(t)$  holds then $|S_t| \le  k/2^t$. 

\begin{lemma}
\label{l:E}
Let $E=E(1) \cup \ldots \cup E(\lambda)$ for $\lambda=1+\log k$. 
Assume that $(4\gamma)^{1/2} <1/4$. Then $\Pr[E] \le 1/3$.
\end{lemma}
\begin{proof}
Let $t'=\lceil t/2 \rceil$. We have
\[ 
\Pr[E(t)] \le  \binom{t}{t'} \gamma^{t'} 
 \le  2^t \gamma^{t'} 
\le  (4\gamma)^{t/2}
\]
Therefore
\[
\Pr[E] \le \sum_t \Pr[E(t)] \le \frac{ (4\gamma)^{1/2}}{1-(4\gamma)^{1/2}} \le 1/4 \cdot 4/3 = 1/3.
\]
\end{proof}

\begin{theorem}
  Suppose $\wh{x}$ is $k$-sparse with entries from $\{-L, \dotsc, L\}$
  for some known $L = n^{O(1)}$.  Then the algorithm
  \textsc{NoiselessSparseFFT} runs in expected $O(k \log n)$ time and
  returns the correct vector $\wh{x}$ with probability at least $2/3$.
\end{theorem}

\begin{proof}
  The correctness follows from Lemma~\ref{l:E}.  The running time is
  dominated by $O(\log k)$ executions of \textsc{HashToBins}.

  Assuming a correct run, in every round $t$ we have
  \[
  \norm{0}{\wh{z}} \leq \norm{0}{\wh{x}} + \abs{S_t} \leq k + k/2^t \leq 2k.
  \]
  Therefore
  \[
  \E[\abs{\{i \in \supp(z) \mid \Eo(i)\}}] \leq \alpha\norm{0}{\wh{z}} \leq 2\alpha k,
  \]
  so the expected running time of each execution of
  \textsc{HashToBins} is $O(\frac{B}{\alpha} \log (n/\delta) + k +
  \alpha k \log (n/\delta))= O(\frac{B}{\alpha} \log n + k + \alpha k
  \log n)$.  Setting $\alpha = \Theta(2^{-t/2})$ and $\beta =
  \Theta(1)$, the expected running time in round $t$ is $O(2^{-t/2}k
  \log n + k + 2^{-t/2} k \log n)$.  Therefore the total expected
  running time is $O(k \log n)$.
\end{proof}

\section{Algorithm for the general case}\label{sec:general}

This section shows how to achieve Equation~\eqref{e:l2l2} for $C = 1 + \eps$.
Pseudocode is in Algorithm~\ref{a:approximate}
and~\ref{a:approximate2}.

\subsection{Intuition}
\label{ss:int}

Let $S$ denote the ``heavy'' $O(k/\eps)$ coordinates of $\wh{x}$.  The
overarching algorithm \textsc{SparseFFT} works by first ``locating'' a
set $L$ containing most of $S$, then ``estimating'' $\wh{x}_L$ to get
$\wh{z}$.  It then repeats on $\wh{x - z}$.  We will show that each
heavy coordinate has a large constant probability of both being in $L$
and being estimated well.  As a result, $\wh{x-z}$ is probably nearly
$k/4$-sparse, so we can run the next iteration with $k \to k/4$.  The
later iterations then run faster and achieve a higher success
probability, so the total running time is dominated by the time in the
first iteration and the total error probability is bounded by a
constant.

In the rest of this intuition, we will discuss the first iteration of
\textsc{SparseFFT} with simplified constants.  In this iteration,
hashes are to $B = O(k/\eps)$ bins and, with $3/4$ probability, we get
$\wh{z}$ so $\wh{x-z}$ is nearly $k/4$-sparse.  The actual algorithm
will involve a parameter $\alpha$ in each iteration, roughly
guaranteeing that with $1-\sqrt{\alpha}$ probability, we get $\wh{z}$
so $\wh{x-z}$ is nearly $\sqrt{\alpha}k$-sparse; the formal guarantee
will be given by Lemma~\ref{lemma:sparsefftoneloop}.  For this
intuition we only consider the first iteration where $\alpha$ is a
constant.

\paragraph{Location.}
As in the noiseless case, to locate the ``heavy'' coordinates we
consider the ``bins'' computed by \textsc{HashToBins} with $P_{\sigma, a,
  b}$.  This roughly corresponds to first permuting the coordinates
according to the (almost) pairwise independent permutation $P_{\sigma,
  a, b}$, partitioning the coordinates into $B = O(k/\eps)$ ``bins''
of $n/B$ consecutive indices, and observing the sum of values in each
bin.  We get that each heavy coordinate $i$ has a large constant
probability that the following two events occur: no other heavy
coordinate lies in the same bin, and only a small (i.e., $O(\eps/k)$)
fraction of the mass from non-heavy coordinates lies in the same bin.
For such a ``well-hashed'' coordinate $i$, we would like to find its
location $\tau = \ps(i) = \sigma(i-b)$ among the $\eps n/k < n/k$
consecutive values that hash to the same bin.  Let
\begin{align}\label{e:theta*}
  \theta^*_j = \frac{2\pi}{n}(j + \sigma b) \pmod{2\pi}.
\end{align}
so $\theta^*_\tau = \frac{2\pi}{n}\sigma i$.  In the noiseless case,
we showed that the difference in phase in the bin using
$P_{\sigma,0,b}$ and using $P_{\sigma, 1, b}$ is $\theta^*_\tau$ plus
a negligible $O(\delta)$ term.  With noise this may not be true;
however, we can say for any $\beta \in [n]$ that the difference in
phase between using $P_{\sigma, a, b}$ and $P_{\sigma, a+\beta, b}$,
as a distribution over uniformly random $a \in [n]$, is $\beta \theta^*_\tau +
\nu$ with (for example) $\E[\nu^2] = 1/100$ (all operations on phases
modulo $2\pi$).  We can only hope to get a constant number of bits
from such a ``measurement''.  So our task is to find $\tau$ within a
region $Q$ of size $n/k$ using $O(\log (n/k))$ ``measurements'' of
this form.

One method for doing so would be to simply do measurements with random
$\beta \in [n]$.  Then each measurement lies within $\pi/4$ of
$\beta\theta^*_\tau$ with at least $1 - \frac{\E[\nu^2]}{\pi^2/16} >
3/4$ probability.  On the other hand, for $j \neq \tau$ and as a
distribution over $\beta$, $\beta(\theta^*_\tau - \theta^*_j)$ is
roughly uniformly distributed around the circle.  As a result, each
measurement is probably more than $\pi/4$ away from $\beta\theta^*_j$.
Hence $O(\log (n/k))$ repetitions suffice to distinguish among the
$n/k$ possibilities for $\tau$.  However, while the number of
measurements is small, it is not clear how to decode in polylog rather
than $\Omega(n/k)$ time.

To solve this, we instead do a $t$-ary search on the location for $t =
\Theta(\log n)$.  At each of $O(\log_t(n/k))$ levels, we split our
current candidate region $Q$ into $t$ consecutive subregions $Q_1,
\dotsc, Q_t$, each of size $w$.  Now, rather than choosing $\beta \in
[n]$, we choose $\beta \in [\frac{n}{16w}, \frac{n}{8w}]$.  By the
upper bound on $\beta$, for each $q \in [t]$ the values $\{\beta
\theta^*_j \mid j \in Q_q\}$ all lie within $\beta w \frac{2\pi}{n}
\leq \pi/4$ of each other on the circle.  On the other hand, if
$\abs{j - \tau} > 16w$, then $\beta(\theta^*_\tau - \theta^*_j)$ will
still be roughly uniformly distributed about the circle.  As a result,
we can check a single candidate element $e_q$ from each subregion: if
$e_q$ is in the same subregion as $\tau$, each measurement usually
agrees in phase; but if $e_q$ is more than $16$ subregions away, each
measurement usually disagrees in phase.  Hence with $O(\log t)$
measurements, we can locate $\tau$ to within $O(1)$ consecutive
subregions with failure probability $1/t^{\Theta(1)}$.  The decoding time is
$O(t \log t)$.

This primitive~\textsc{LocateInner} lets us narrow down the candidate
region for $\tau$ to a subregion that is a $t' = \Omega(t)$ factor
smaller.  By repeating \textsc{LocateInner} $\log_{t'} (n/k)$ times,
\textsc{LocateSignal} can find $\tau$ precisely.  The number of
measurements is then $O(\log t \log_t (n/k)) = O(\log (n/k))$ and the
decoding time is $O(t \log t \log_t (n/k)) = O(\log (n/k) \log n)$.
Furthermore, the ``measurements'' (which are actually calls to
\textsc{HashToBins}) are non-adaptive, so we can perform them in
parallel for all $O(k/\eps)$ bins, with $O(\log (n/\delta))$ average
time per measurement.  This gives $O(k \log (n/k) \log (n/\delta))$
total time for \textsc{LocateSignal}.

This lets us locate every heavy and ``well-hashed'' coordinate with
$1/t^{\Theta(1)} = o(1)$ failure probability, so every heavy coordinate is located
with arbitrarily high constant probability.

\paragraph{Estimation.}
By contrast, estimation is fairly simple.  As in
Algorithm~\ref{a:exact}, we can estimate $\wh{(x-z)}_i$ as
$\wh{u}_{\hs(i)}$, where $\wh{u}$ is the output of
\textsc{HashToBins}.  Unlike in Algorithm~\ref{a:exact}, we now have
noise that may cause a single such estimate to be poor even if $i$ is
``well-hashed''.  However, we can show that for a random permutation
$P_{\sigma,a,b}$ the estimate is ``good'' with constant probability.
\textsc{EstimateValues} takes the median of $R_{est} = O(\log
\frac{1}{\eps})$ such samples, getting a good estimate with
$1-\eps/64$ probability.  Given a candidate set $L$ of size $k/\eps$,
with $7/8$ probability at most $k/8$ of the coordinates are badly
estimated.  On the other hand, with $7/8$ probability, at least $7k/8$
of the heavy coordinates are both located and well estimated.  This
suffices to show that, with $3/4$ probability, the largest $k$
elements $J$ of our estimate $\wh{w}$ contains good estimates of
$3k/4$ large coordinates, so $\wh{x-z-w_J}$ is close to $k/4$-sparse.

\begin{algorithm}[btp]
  \begin{algorithmic}
    \Procedure{SparseFFT}{$x$, $k$, $\eps$, $\delta$} 
    \State $R \gets O(\log k / \log \log k)$ as in Theorem~\ref{thm:sparsefft}.
    \State $\wh{z}^{(1)} \gets 0$
    \For{$r \in [R]$}
    \State Choose $B_r, k_r, \alpha_r$ as in Theorem~\ref{thm:sparsefft}.
    \State $R_{est} \gets O(\log(\frac{B_r}{\alpha_rk_r}))$ as in Lemma~\ref{lemma:sparsefftoneloop}.
    \State $L_r \gets \textsc{LocateSignal}(x, \wh{z}^{(r)}, B_r, \alpha_r, \delta)$
    \State $\wh{z}^{(r+1)} \gets \wh{z}^{(r)} + \textsc{EstimateValues}(x, \wh{z}^{(r)}, 3k_r, L_r, B_r, \delta, R_{est})$. 
    \EndFor
    \State \Return $\wh{z}^{(R+1)}$
    \EndProcedure
    \Procedure{EstimateValues}{$x$, $\wh{z}$, $k'$, $L$, $B$, $\delta$, $R_{est}$}
    \For{$r \in [R_{est}]$}
      \State Choose  $a_r, b_r \in [n]$ uniformly at random.
      \State Choose $\sigma_r$ uniformly at random from the set of odd numbers in $[n]$.
      \State $\wh{u}^{(r)} \gets \textsc{HashToBins}(x, \wh{z}, P_{\sigma,a_r,b}, B, \delta)$.
    \EndFor
    \State $\wh{w}\gets 0$ 
    \For{$i \in L$}
     \State $\wh{w}_{i} \gets \median_{r} \wh{u}_{\hs(i)}^{(r)} \omega^{-a_r \sigma i}$.\Comment{Separate median in real and imaginary axes.}
    \EndFor
    \State $J \gets \argmax_{\abs{J} = k'} \norm{2}{\wh{w}_J}$.
    \State \Return $\wh{w}_J$
    \EndProcedure
  \end{algorithmic}
  \caption{$k$-sparse recovery for general signals, part 1/2.}\label{a:approximate}
\end{algorithm}

\begin{algorithm}[hp]
  \begin{algorithmic}
    \Procedure{LocateSignal}{$x$, $\wh{z}$, $B$, $\alpha$, $\delta$}

    \State Choose uniformly at random $\sigma, b \in [n]$ with $\sigma$ odd.
    \State Initialize $l^{(1)}_i = (i-1)n/B$ for $i \in [B]$.
    \State Let $w_0 = n/B, t = \log n, t' = t/4, D_{max}=\log_{t'} (w_0+1)$.
    \State Let $R_{loc} = \Theta(\log_{1/\alpha} (t/\alpha))$ per Lemma~\ref{lemma:looplocation}.
    \For{$D \in [D_{max}]$}
      \State $l^{(D+1)} \gets \textsc{LocateInner}(x, \wh{z}, B, \delta, \alpha, \sigma, \beta, l^{(D)},  w_0/(t')^{D-1}, t,  R_{loc})$
    \EndFor
    \State $L \gets \{\ps^{-1}(l^{(D_{max}+1)}_j) \mid j \in [B]\}$
    \State \Return $L$
    \EndProcedure
    \State
    \Comment{$\delta, \alpha$ parameters for $G$, $G'$}\\
    \Comment{$(l_1, l_1+w), \dotsc, (l_B, l_B+w)$ the plausible regions.}\\
    \Comment{$B \approx k/\eps$ the number of bins}\\
    \Comment{$t \approx \log n$ the number of regions to split into.}\\
    \Comment{$R_{loc} \approx \log t = \log \log n$ the number of rounds to run}
    \Procedure{LocateInner}{$x$, $\wh{z}$, $B$, $\delta$, $\alpha$, $\sigma$, $b$, $l$, $w$, $t$, $R_{loc}$}
    \State Let $s = \Theta(\alpha^{1/3})$.
    \State Let $v_{j,q} = 0$ for $(j,q) \in [B]\times[t]$.
    \For{$r \in [R_{loc}]$}
     \State Choose $a \in [n]$ uniformly at random.
     \State Choose $\beta \in \{\frac{snt}{4w}, \dotsc, \frac{snt}{2w}\}$ uniformly at random.
     \State $\wh{u} \gets \textsc{HashToBins}(x, \wh{z}, P_{\sigma,a,b}, B, \delta, \alpha)$.
     \State $\wh{u}' \gets \textsc{HashToBins}(x, \wh{z}, P_{\sigma,a+\beta,b}, B, \delta, \alpha)$.
      \For{$j \in [B]$}
       \State $c_j \gets \phi(\wh{u}_j / \wh{u}'_j)$
       \For{$q \in [t]$}
        \State $m_{j,q} \gets l_j + \frac{q - 1/2}{t}w$
        \State $\theta_{j,q} \gets \frac{2\pi  (m_{j,q} + \sigma b)}{n} \bmod 2\pi$
        \If{$\min(\abs{\beta\theta_{j,q} -c_j}, 2\pi-\abs{\beta\theta_{j,q}-c_j}) < s\pi$}
         \State $v_{j,q} \gets v_{j,q} + 1$
        \EndIf
       \EndFor
      \EndFor
    \EndFor
    \For{$j \in [B]$}
      \State $Q^* \gets \{q \in [t] \mid v_{j,q} > R_{loc}/2\}$
      \If{$Q^* \neq \emptyset$}
      \State $l_j' \gets \min_{q \in Q^*} l_j + \frac{q-1}{t}w$
      \Else
      \State $l_j' \gets \perp$
      \EndIf
    \EndFor
    \State \Return $l'$
    \EndProcedure
  \end{algorithmic}
  \caption{$k$-sparse recovery for general signals, part 2/2.}\label{a:approximate2}
\end{algorithm}

\subsection{Formal definitions}\label{ss:genformal}

As in the noiseless case, we define $\ps(i) = \sigma (i-b) \bmod n$,
$\hs(i) = \text{round}(\ps(i) B/n)$ and $\os(i)=\ps(i) - \hs(i) n/B$.
We say $\hs(i)$ is the ``bin'' that frequency $i$ is mapped into, and
$\os(i)$ is the ``offset''.  We define $\hs^{-1}(j) = \{i \in [n]
\mid \hs(i) = j\}$.

Define
\[
\err(x, k) = \min_{k\text{-sparse\ } y} \norm{2}{x-y}.
\]
In each iteration of \textsc{SparseFFT}, define $\wh{x}' =
\wh{x}-\wh{z}$, and let
\begin{align*}
  \rho^2 &= \err^2(\wh{x'}, k) + \delta^2n\norm{1}{\wh{x}}^2\\
  \mu^2 &= \eps\rho^2/k\\
  S &= \{i \in [n] \mid \sabs{\wh{x'}_i}^2 \geq \mu^2\}
\end{align*}
Then $\abs{S} \leq (1 + 1/\eps)k = O(k/\eps)$ and $\norm{2}{\wh{x'}-\wh{x'}_S}^2
\leq (1+\eps)\rho^2$.  We will show that each $i \in S$ is found by
\textsc{LocateSignal} with probability $1-O(\alpha)$, when $B =
\Omega(\frac{k}{\alpha\eps})$.

For any $i \in S$ define three types of events associated with $i$ and
$S$ and defined over the probability space induced by $\sigma$ and
$b$:
\begin{Itemize}
\item ``Collision'' event $\Ec(i)$: holds iff $\hs(i) \in \hs(S\setminus\{i\})$;
\item ``Large offset'' event $\Eo(i)$: holds iff $\sabs{\os(i)} \ge (1-\alpha) n/(2B)$; and
\item ``Large noise'' event $\En(i)$: holds iff
  $\norm{2}{\wh{x'}_{\hs^{-1}(\hs(i)) \setminus S}}^2 \geq \err^2(\wh{x'}, k) /
  (\alpha B)$.
\end{Itemize}

By Claims~\ref{c:coll} and~\ref{c:off}, $\Pr[\Ec(i)] \leq 4\abs{S}/B =
O(\alpha)$ and $\Pr[\Eo(i)] \leq 2\alpha$ for any $i \in S$.

\begin{claim}\label{c:noise}
  For any $i \in S$, $\Pr[\En(i)] \leq 4\alpha$.
\end{claim}
\begin{proof}
  For each $j \neq i$, $\Pr[\hs(j) = \hs(i)] \leq \Pr[\abs{\sigma j -
    \sigma i} < n/B] \leq 4/B$ by
  Lemma~\ref{lemma:limitedindependence}.  Then
  \[
  \E[\norm{2}{\wh{x'}_{\hs^{-1}(\hs(i)) \setminus S}}^2] \leq 4\norm{2}{\wh{x'}_{[n]\setminus S}}^2/B
  \]
  The result follows by Markov's inequality.
\end{proof}

We will show for $i \in S$ that if none of $\Ec(i), \Eo(i)$, and
$\En(i)$ hold then \textsc{SparseFFTInner} recovers $\wh{x}'_i$ with
$1 - O(\alpha)$ probability.

\begin{lemma}\label{lemma:hashingnoise}
  Let $a \in [n]$ uniformly at random, $B$ divide $n$, and the other
  parameters be arbitrary in
  \[
  \wh{u} = \textsc{HashToBins}(x, \wh{z}, P_{\sigma, a, b}, B, \delta, \alpha).
  \]
  Then for any $i \in [n]$ with $j = \hs(i)$ and none of $\Ec(i)$,
  $\Eo(i)$, or $\En(i)$ holding,
  \[
  \E[ \abs{\wh{u}_j - \wh{x'}_i\omega^{a\sigma i}}^2] \leq 2 \frac{\rho^2}{\alpha B}
  \]
\end{lemma}
\begin{proof}
  Let $\wh{G'} = \wh{G'}_{B,\delta,\alpha}$.  Let $T = \hs^{-1}(j)
  \setminus \{i\}$.  We have that $T \cap S = \{\}$ and
  $\wh{G'}_{-\os(i)} = 1$.  By Lemma~\ref{lemma:hashtobins},
  \begin{align*}
    \wh{u}_j - \wh{x'}_{i}\omega^{a\sigma i} &= \sum_{i' \in T} \wh{G'}_{-o_\sigma(i')}\wh{x'}_{i'} \omega^{a\sigma i'} \pm
    \delta\norm{1}{\wh{x}}.
  \end{align*}
  Because the $\sigma i'$ are distinct for $i' \in T$, we have by Parseval's theorem
  \begin{align*}
    \E_a \left|\sum_{i' \in T} \wh{G'}_{-o_\sigma(i')}\wh{x'}_{i'} \omega^{a\sigma i'}\right|^2
    &= \sum_{i' \in T} (\wh{G'}_{-o_\sigma(i')}\wh{x'}_{i'})^2 \leq \norm{2}{\wh{x'_T}}^2
  \end{align*}
  Since $\abs{X+Y}^2 \leq 2\abs{X}^2 + 2\abs{Y}^2$ for any $X, Y$, we get
  \begin{align*}
    \E_a[\abs{\wh{u}_j - \wh{x'}_{i}\omega^{a\sigma i}}^2] &\leq 2\norm{2}{x'_T}^2 + 2\delta^2\norm{1}{\wh{x}}^2\\
    &\leq 2\err^2(\wh{x'}, k)/(\alpha B) + 2\delta^2\norm{1}{\wh{x}}^2\\
    &\leq 2 \rho^2 / (\alpha B).
  \end{align*}
\end{proof}

\subsection{Properties of \textsc{LocateSignal}}

In our intuition, we made a claim that if $\beta \in [n/(16 w),
n/(8w)]$ uniformly at random, and $i > 16w$, then $\frac{2\pi}{n}\beta
i$ is ``roughly uniformly distributed about the circle'' and hence not
concentrated in any small region.  This is clear if $\beta$ is chosen
as a random real number; it is less clear in our setting where $\beta$
is a random integer in this range.  We now prove a lemma that
formalizes this claim.

\begin{lemma}\label{lemma:rectangle}
  Let $T \subset [m]$ consist of $t$ consecutive integers, and suppose
  $\beta \in T$ uniformly at random.  Then for any $i \in [n]$ and
  set $S \subset [n]$ of $l$ consecutive integers,
  \[
  \Pr[\beta i \bmod n \in S] \leq \ceil{im/n}(1 + \floor{l/i})/t \leq
  \frac{1}{t} + \frac{im}{nt} + \frac{lm}{nt} + \frac{l}{it}
  \]
\end{lemma}
\begin{proof}
  Note that any interval of length $l$ can cover at most $1 +
  \floor{l/i}$ elements of any arithmetic sequence of common
  difference $i$.  Then $\{\beta i \mid \beta \in T\} \subset [im]$
  is such a sequence, and there are at most $\ceil{im/n}$ intervals
  $an + S$ overlapping this sequence.  Hence at most $\ceil{im/n}(1 +
  \floor{l/i})$ of the $\beta \in [m]$ have $\beta i \bmod n \in S$.
  Hence
  \[
  \Pr[\beta i \bmod n \in S] \leq \ceil{im/n}(1 + \floor{l/i})/t.
  \]
\end{proof}

\begin{lemma}\label{lemma:LocateInner}
  Let $i \in S$.  Suppose none of $\Ec(i), \Eo(i)$, and $\En(i)$ hold,
  and let $j = \hs(i)$.  Consider any run of \textsc{LocateInner} with
  $\ps(i) \in [l_j, l_j+w]$ .  Let $f > 0$ be a parameter such that
  \[
  B = \frac{Ck}{\alpha f \eps}.
  \]
  for $C$ larger than some fixed constant.  Then $\ps(i) \in [l'_j,
  l'_j + 4w/t]$ with probability at least $1-tf^{\Omega(R_{loc})}$,
\end{lemma}
\begin{proof}
  Let $\tau = \ps(i) \equiv \sigma (i - b) \pmod{n}$, and for any
  $j \in [n]$ define
  \[
  \theta^*_j = \frac{2\pi}{n}(j + \sigma b) \pmod{2\pi}
  \]
  so $\theta^*_\tau = \frac{2\pi}{n}\sigma i$.  Let $g =
  \Theta(f^{1/3})$, and $C' = \frac{B\alpha\eps}{k} = \Theta(1/g^3)$.

  To get the result, we divide $[l_j, l_j+w]$ into $t$ ``regions'',
  $Q_q = [l_j + \frac{q-1}{t}w, l_j + \frac{q}{t}w]$ for $q \in [t]$.
  We will first show that in each round $r$, $c_j$ is close to $\beta
  \theta^*_\tau$ with $1-g$ probability.  This will imply that $Q_q$
  gets a ``vote,'' meaning $v_{j,q}$ increases, with $1-g$ probability
  for the $q'$ with $\tau \in Q_{q'}$.  It will also imply that
  $v_{j,q}$ increases with only $g$ probability when $\abs{q - q'} >
  3$.  Then $R_{loc}$ rounds will suffice to separate the two with $1
  - f^{-\Omega(R_{loc})}$ probability.  We get that with $1 -
  tf^{-\Omega(R_{loc})}$ probability, the recovered $Q^*$ has
  $\abs{q-q'} \leq 3$ for all $q \in Q^*$.  If we take the minimum $q
  \in Q^*$ and the next three subregions, we find $\tau$ to within $4$
  regions, or $4w/t$ locations, as desired.

  In any round $r$, define $\wh{u} = \wh{u}^{(r)}$ and $a = a_r$.  We
  have by Lemma~\ref{lemma:hashingnoise} and that $i \in S$ that
  \begin{align*}
    \E[\abs{\wh{u}_j - \omega^{a\sigma i}\wh{x'}_i}^2] &\leq 2\frac{\rho^2}{\alpha B} = \frac{2k}{B\alpha \eps} \mu^2\\
    &= \frac{2}{C'}\mu^2 \leq \frac{2}{C'}\sabs{\wh{x'}_i}^2.
  \end{align*}
  Note that $\phi(\omega^{a\sigma i}) = -a\theta^*_\tau$.  Thus for any $p
  > 0$, with probability $1-p$ we have
  \begin{align*}
    \abs{\wh{u}_j - \omega^{a\sigma i}\wh{x'}_i} &\leq \sqrt{\frac{2}{C' p}}\abs{\wh{x'}_i}\\
  \norm{\bigcirc}{\phi(\wh{u}_j) - (\phi(\wh{x'}_i) - a\theta^*_\tau)} &\leq \sin^{-1}(\sqrt{\frac{2}{C' p}})
  \end{align*}
  where $\norm{\bigcirc}{x-y} = \min_{\gamma \in \Z}\abs{x-y + 2\pi
    \gamma}$ denotes the ``circular distance'' between $x$ and $y$.
  The analogous fact holds for $\phi(\wh{u'}_j)$ relative to
  $\phi(\wh{x'}_i) - (a+\beta)\theta^*_\tau$. Therefore with at least
  $1-2p$ probability,
  \begin{align*}
        \norm{\bigcirc}{c_j - \beta \theta^*_\tau} &= \norm{\bigcirc}{\phi(\wh{u}_j) - \phi(\wh{u'}_j) - \beta \theta^*_\tau}\\
    &= \bigg\|\left(\phi(\wh{u}_j) - (\phi(\wh{x'}_i) - a\theta^*_\tau)\right) -
      \left(\phi(\wh{u'}_j) - (\phi(\wh{x'}_i) - (a+\beta)\theta^*_\tau)\right)\bigg\|_\bigcirc\\
     &\leq \norm{\bigcirc}{\phi(\wh{u}_j) - (\phi(\wh{x'}_i) - a\theta^*_\tau)} + 
     \norm{\bigcirc}{\phi(\wh{u'}_j) - (\phi(\wh{x'}_i) - (a+\beta)\theta^*_\tau)}\\
     &\leq 2\sin^{-1}(\sqrt{\frac{2}{C' p}})
  \end{align*}
  by the triangle inequality.  Thus for any $s = \Theta(g)$ and $p =
  \Theta(g)$, we can set $C' = \frac{2}{p\sin^2(s\pi/4)} = \Theta(1/g^3)$ so
  that
  \begin{align}\label{eq:cj}
    \norm{\bigcirc}{c_j - \beta\theta^*_\tau} < s\pi/2
  \end{align}
  with probability at least $1-2p$.

  Equation~\eqref{eq:cj} shows that $c_j$ is a good estimate for $i$
  with good probability.  We will now show that this means the
  approprate ``region'' $Q_{q'}$ gets a ``vote'' with ``large''
  probability.

  For the $q'$ with $\tau \in [l_j + \frac{q'-1}{t}w, l_j +
  \frac{q'}{t}w]$, we have that $m_{j,q'} = l_j + \frac{q'-1/2}{t}w$
  satisfies
  \[
  \abs{\tau - m_{j,q'}} \leq \frac{w}{2t}
  \]
  so
  \[
  \abs{\theta^*_\tau - \theta_{j,q'}} \leq \frac{2\pi}{n}\frac{w}{2t}.
  \]
  Hence by Equation~\eqref{eq:cj}, the triangle inequality, and the
  choice of $B \leq \frac{snt}{2w}$,
  \begin{align*}
    \norm{\bigcirc}{c_j - \beta\theta_{j,q'}} &\leq
    \norm{\bigcirc}{c_j - \beta \theta^*_\tau}
    + \norm{\bigcirc}{\beta \theta^*_\tau - \beta \theta_{j,q'}}\\
    &< \frac{s\pi}{2} + \frac{\beta \pi w}{n t}\\
    &\leq \frac{s\pi}{2} + \frac{s\pi}{2}\\
    & = s \pi.
  \end{align*}
  Thus, $v_{j,q'}$ will increase in each round with probability at
  least $1-2p$.

  Now, consider $q$ with $\abs{q-q'} > 3$.  Then $\abs{\tau - m_{j,
      q}} \geq \frac{7w}{2t}$, and (from the definition of $\beta >
  \frac{snt}{4w}$) we have
  \begin{eqnarray}
    \label{e:beta}
    \beta\abs{\tau - m_{j,q}} \geq \frac{7sn}{8} > \frac{3sn}{4}.
  \end{eqnarray}

  We now consider two cases. First, suppose that $ \abs{\tau - m_{j,q}}
  \leq \frac{w}{st}$.  In this case, from the definition of $\beta$ it
  follows that
  \[
  \beta\abs{\tau - m_{j,q}} \leq n/2.
  \]

  Together with Equation~\eqref{e:beta} this implies
  \[
  \Pr[\beta (\tau - m_{j,q}) \bmod n \in [-3s n/4, 3sn/4]] = 0.
  \]

  On the other hand, suppose that $\abs{\tau - m_{j,q}} >
  \frac{w}{st}$.  In this case, we use Lemma~\ref{lemma:rectangle}
  with parameters $l=3sn/2$, $m=\frac{snt}{2w}$, $t=\frac{snt}{4w}$,
  $i=(\tau - m_{j,q})$ and $n=n$, to conclude that
  \begin{align*}
    \Pr[\beta (\tau - m_{j,q}) \bmod n \in [-3s n/4, 3sn/4]]
    &\leq
    \frac{4w}{snt} + 2\frac{\abs{\tau - m_{j,q}}}{n} + 3s +
    \frac{3sn}{2}\frac{st}{w}\frac{4w}{snt}\\
    &\leq \frac{4w}{snt} + \frac{2w}{n} + 9s\\
    &< \frac{6}{sB} + 9s   \\
    &< 10s
  \end{align*}
  where we used that $\abs{i} \leq w \leq n/B$, the assumption
  $\frac{w}{st} < |i|$, $t \geq 1$, $s < 1$, and that $s^2 > 6/B$
  (because $s = \Theta(g)$ and $B = \omega(1/g^3)$).

  Thus in either case, with probability at least $1 - 10s$ we have
  \begin{align*}
    \norm{\bigcirc}{\beta\theta_{j,q} - \beta \theta^*_\tau} =  \norm{\bigcirc}{\frac{2\pi \beta (m_{j,q}-\tau)}{n}} > \frac{2\pi}{n}\frac{3sn}{4} = \frac{3}{2}s \pi
  \end{align*}
  for any $q$ with $\abs{q-q'} > 3$.  Therefore
we have
  \[
  \norm{\bigcirc}{c_j - \beta\theta_{j, q}} \geq \norm{\bigcirc}{\beta\theta_{j,q} - \beta \theta^*_\tau} - \norm{\bigcirc}{c_j - \beta \theta^*_\tau} >
  s \pi
  \]
  with probability at least $1 - 10s - 2p$, and $v_{j, q}$ is not
  incremented.

  To summarize: in each round, $v_{j,q'}$ is incremented with
  probability at least $1 - 2p$ and $v_{j, q}$ is incremented with
  probability at most $10s + 2p$ for $\abs{q-q'} > 3$. The
  probabilities corresponding to different rounds are independent.

  Set $s = g/20$ and $p = g/4$.  Then $v_{j,q'}$ is incremented with
  probability at least $1-g$ and $v_{j,q}$ is incremented with
  probability less than $g$.  Then after $R_{loc}$ rounds, if
  $\abs{q-q'} > 3$,
  \[
  \Pr[v_{j,q} > R_{loc}/2] \leq \binom{R_{loc}}{R_{loc}/2} g^{R_{loc}/2} \leq (4g)^{R_{loc}/2} = f^{\Omega(R_{loc})}
  \]
  for $g = f^{1/3}/4$.  Similarly,
  \[
  \Pr[v_{j,q'} < R_{loc}/2] \leq f^{\Omega(R_{loc})}.
  \]
  Hence with probability at least $1 - tf^{\Omega(R_{loc})}$ we have $q'
  \in Q^*$ and $\abs{q-q'} \leq 3$ for all $q \in Q^{*}$.  But then
  $\tau - l'_j \in [0, 4w/t]$ as desired.

  Because $\E[\abs{\{i \in \supp(\wh{z})\mid \Eo(i)\}}] = \alpha
  \norm{0}{\wh{z}}$, the expected running time is $O(R_{loc}Bt +
  R_{loc}\frac{B}{\alpha}\log (n/\delta) + R_{loc} \norm{0}{\wh{z}}
  (1 + \alpha \log (n/\delta)))$.
\end{proof}

\begin{lemma}\label{lemma:looplocation}
  Suppose $B = \frac{Ck}{\alpha^2\eps}$ for $C$ larger than some fixed
  constant.  The procedure \textsc{LocateSignal} returns a set $L$ of
  size $\abs{L} \leq B$ such that for any $i \in S$, $\Pr[i \in L]
  \geq 1 - O(\alpha)$. Moreover the procedure runs in expected time
  \[
  O((\frac{B}{\alpha}\log (n/\delta) + \norm{0}{\wh{z}}(1 + \alpha \log (n/\delta))) \log (n/B)).
  \]
\end{lemma}
\begin{proof}
  Consider any $i \in S$ such that none of $\Ec(i), \Eo(i)$, and
  $\En(i)$ hold, as happens with probability $1 - O(\alpha)$.

  Set $t = \log n, t' = t/4$ and $R_{loc} = O(\log_{1/\alpha}
  (t/\alpha))$.  Let $w_0 = n/B$ and $w_D = w_0/(t')^{D-1}$, so
  $w_{D_{max}+1} < 1$ for $D_{max} = \log_{t'} (w_0+1) < t$.  In each
  round $D$, Lemma~\ref{lemma:LocateInner} implies that if $\ps(i) \in
  [l^{(D)}_j, l^{(D)}_j+w_D]$ then $\ps(i) \in [l^{(D+1)}_j,
  l^{(D+1)}_j+w_{D+1}]$ with probability at least $1 -
  \alpha^{\Omega(R_{loc})} = 1 - \alpha / t$.  By a union bound, with
  probability at least $1 - \alpha$ we have $\ps(i) \in
  [l^{(D_{max}+1)}_j, l^{(D_{max}+1)}_j + w_{D_{max}+1}] =
  \{l^{(D_{max}+1)}_j\}$.  Thus $i = \ps^{-1}(l^{(D_{max}+1)}_j) \in
  L$.

  Since $R_{loc}D_{max} = O(\log_{1/\alpha}(t/\alpha) \log_t (n/B)) =
  O(\log (n/B))$, the running time is
  \begin{align*}
    &
    O(D_{max}(R_{loc} \frac{B}{\alpha}\log (n/\delta) + R_{loc}
    \norm{0}{\wh{z}}(1 + \alpha \log (n/\delta)))) 
    \\
    ={}&
    O((\frac{B}{\alpha}\log (n/\delta) + \norm{0}{\wh{z}}(1 +
    \alpha \log (n/\delta)))\log (n/B)).
  \end{align*}
\end{proof}


\subsection{Properties of \textsc{EstimateValues}}

\begin{lemma}\label{lemma:indexestimate}
  For any $i \in L$,
  \[
  \Pr[\abs{\wh{w}_i - \wh{x'}_i}^2 > \mu^2] < e^{-\Omega(R_{est})}
  \]
  if $B > \frac{Ck}{\alpha\eps}$ for some constant $C$.
\end{lemma}
\begin{proof}
  Define $e_r = \wh{u}_j^{(r)}\omega^{-a_r \sigma i} - \wh{x'}_i$ in each round
  $r$.  Suppose none of $\Ec^{(r)}(i), \Eo^{(r)}(i)$, and
  $\En^{(r)}(i)$ hold, as happens with probability $1 - O(\alpha)$.
  Then by Lemma~\ref{lemma:hashingnoise},
  \begin{align*}
    \E_{a_r}[\abs{e_r}^2] &\leq 2\frac{\rho^2}{\alpha B} = \frac{2k}{\alpha \eps B}\mu^2 < \frac{2}{C}\mu^2
  \end{align*}
  Hence with $3/4 - O(\alpha) > 5/8$ probability in total,
  \[
  \abs{e_r}^2 < \frac{8}{C} \mu^2 < \mu^2/2
  \]
  for sufficiently large $C$.  Then with probability at least $1 -
  e^{-\Omega(R_{est})}$, both of the following occur:
  \begin{align*}
    \abs{\median_r \text{real}( e_r)}^2 &< \mu^2/2\\
    \abs{\median_r \text{imag}(e_r)}^2 &< \mu^2/2.
  \end{align*}
  If this is the case, then $\abs{\median_r e_r}^2 < \mu^2$.  Since $\wh{w}_i = \wh{x'}_i + \median e_r$,
  the result follows.
\end{proof}

\begin{lemma}\label{lemma:estimation}
  Let $R_{est} \geq C\log \frac{B}{\gamma f k}$ for some constant $C$
  and parameters $\gamma, f > 0$.  Then if \textsc{EstimateValues} is
  run with input $k' = 3k$, it returns $\wh{w_J}$ for $\abs{J} = 3k$
  satisfying
  \[
  \Err^2(\wh{x'_L} - \wh{w_J}, fk) \leq \Err^2(\wh{x'_L}, k) + O(k \mu^2)
  \]
  with probability at least $1 - \gamma$.
\end{lemma}
\begin{proof}
  By Lemma~\ref{lemma:indexestimate}, each index $i \in L$ has
  \[
  \Pr[\abs{\wh{w}_i - \wh{x'}_i}^2 > \mu^2] < \frac{\gamma f k}{B}.
  \]
  Let $U = \{i \in L \mid \abs{\wh{w}_i - \wh{x'}_i}^2 > \mu^2\}$.  With
  probability $1 - \gamma$, $\abs{U} \leq fk$; assume this happens.
  Then
  \begin{align}
    \label{e:infty}
    \norm{\infty}{(\wh{x'} - \wh{w})_{L \setminus U}}^2 \leq \mu^2.
  \end{align}
  Let $T$ contain the top $2k$ coordinates of $\wh{w}_{L \setminus U}$.
  By the analysis of Count-Sketch (most specifically, Theorem~3.1
  of~\cite{PW}), the $\ell_\infty$ guarantee~\eqref{e:infty} means that
  \begin{eqnarray}
  \label{e:l2}
  \norm{2}{\wh{x'}_{L \setminus U} - \wh{w}_T}^2 \leq \Err^2(\wh{x'}_{L \setminus U}, k) + 3k\mu^2.
  \end{eqnarray}
  Because $J$ is the top $3k > (2+f)k$ coordinates of $\wh{w_L}$, $T
  \subset J$.  Let $J' = J \setminus (T \cup U)$, so $\abs{J'} \leq
  k$.  Then
  \begin{align*}
    \Err^2(\wh{x'_L} - \wh{w_J}, fk) &\leq
    \norm{2}{\wh{x'_{L\setminus U}} - \wh{w_{J\setminus U}}}^2\\
    &= \norm{2}{\wh{x'}_{L\setminus (U \cup J')} - \wh{w_{T}}}^2 + \norm{2}{(\wh{x'}-\wh{w})_{J'}}^2\\
    &\leq \norm{2}{\wh{x'}_{L\setminus U} - \wh{w_{T}}}^2 + \abs{J'}\norm{\infty}{(\wh{x'}-\wh{w})_{J'}}^2\\
    &\leq \Err^2(\wh{x'}_{L \setminus U}, k) + 3k\mu^2 + k\mu^2\\
    &= \Err^2(\wh{x'}_{L \setminus U}, k) + O(k \mu^2)
  \end{align*}
where we used Equations~\eqref{e:infty} and~\eqref{e:l2}.
\end{proof}

\subsection{Properties of \textsc{SparseFFT}}

We will show that $\wh{x}-\wh{z}^{(r)}$ gets sparser as $r$ increases, with
only a mild increase in the error.

\begin{lemma}\label{lemma:sparsefftoneloop}
  Define $\wh{x}^{(r)} = \wh{x}-\wh{z}^{(r)}$.  Consider any one loop
  $r$ of \textsc{SparseFFT}, running with parameters $(B, k, \alpha) = (B_r, k_r, \alpha_r)$
  such that $B \geq \frac{Ck}{\alpha^2 \eps}$ for some $C$ larger than
  some fixed constant.  Then for any $f > 0$,
  \[
  \Err^2(\wh{x}^{(r+1)}, fk) \leq (1+\eps)\Err^2(\wh{x}^{(r)}, k) + O(\eps\delta^2n\norm{1}{\wh{x}}^2)
  \]
  with probability $1 - O(\alpha/f)$, and the running time is
  \[
  O((\snorm{0}{\wh{z}^{(r)}}(1 + \alpha \log (n/\delta)) + \frac{B}{\alpha} \log (n/\delta)) (\log \frac{1}{\alpha\eps} + \log (n/B))).
  \]
\end{lemma}
\begin{proof}
  We use $R_{est} = O(\log \frac{B}{\alpha k}) = O(\log \frac{1}{\alpha\eps})$
  rounds inside \textsc{EstimateValues}.

  The running time for \textsc{LocateSignal} is
  \[
  O((\frac{B}{\alpha}\log (n/\delta) + \snorm{0}{\wh{z}^{(r)}}(1 + \alpha \log (n/\delta))) \log (n/B)),
  \]
  and for \textsc{EstimateValues} is
  \[
  O((\frac{B}{\alpha}\log (n/\delta) + \snorm{0}{\wh{z}^{(r)}}(1 + \alpha \log (n/\delta)))\log \frac{1}{\alpha\eps} )
  \]
  for a total running time as given.

  Recall that in round $r$, $\mu^2 =
  \frac{\eps}{k}(\Err^2(\wh{x}^{(r)}, k) + \delta^2 n
  \norm{1}{\wh{x}}^2)$ and $S = \{i \in [n] \mid
  \abs{\wh{x}^{(r)}_i}^2 > \mu^2\}$.  By
  Lemma~\ref{lemma:looplocation}, each $i \in S$ lies in $L_r$ with
  probability at least $1 - O(\alpha)$.  Hence $\abs{S \setminus L} <
  fk$ with probability at least $1 - O(\alpha/f)$.  Then
  \begin{align}\label{e:othererr}
  \Err^2(\wh{x}^{(r)}_{[n] \setminus L}, fk) &\leq \norm{2}{\wh{x}^{(r)}_{[n]
    \setminus (L \cup S)}}^2\notag\\
  &\leq \err^2(\wh{x}^{(r)}_{[n] \setminus (L \cup S)}, k) + k \norm{\infty}{\wh{x}^{(r)}_{[n] \setminus (L \cup S)}}^2
  \notag\\&
  \leq \err^2(\wh{x}^{(r)}_{[n] \setminus L}, k) + k \mu^2.
  \end{align}
  Let $\wh{w} = \wh{z}^{(r+1)}-\wh{z}^{(r)} = \wh{x}^{(r)} -
  \wh{x}^{(r+1)}$ by the vector recovered by \textsc{EstimateValues}.
  Then $\supp(\wh{w}) \subset L$, so
  \begin{align*}
    \Err^2(\wh{x}^{(r+1)}, 2fk) &= \Err^2(\wh{x}^{(r)} - \wh{w}, 2fk)\\
    &\leq \Err^2(\wh{x}^{(r)}_{[n] \setminus L}, fk) + \Err^2(\wh{x}^{(r)}_L-\wh{w}, fk)\\
    &\leq \Err^2(\wh{x}^{(r)}_{[n] \setminus L}, fk) + \Err^2(\wh{x}^{(r)}_L, k) + O(k\mu^2)
  \end{align*}
  by Lemma~\ref{lemma:estimation}.  But by Equation~\eqref{e:othererr}, this gives
  \begin{align*}
    \Err^2(\wh{x}^{(r+1)}, 2fk) &\leq \Err^2(\wh{x}^{(r)}_{[n] \setminus L}, k) + \Err^2(\wh{x}^{(r)}_L, k) + O(k\mu^2)\\
    &\leq \Err^2(\wh{x}^{(r)}, k) + O(k \mu^2)\\
    &=(1 + O(\eps))\Err^2(\wh{x}^{(r)}, k) + O(\eps \delta^2n \norm{1}{\wh{x}}^2).
  \end{align*}
  The result follows from rescaling $f$ and $\eps$ by constant factors.
\end{proof}

Given the above, this next proof follows a similar argument
to~\cite{IPW}, Theorem 3.7.
\begin{theorem}\label{thm:sparsefft}
  With $2/3$ probability, \textsc{SparseFFT} recovers $\wh{z}^{(R+1)}$
  such that
  \[
  \norm{2}{\wh{x}-\wh{z}^{(R+1)}} \leq (1 + \eps) \err(\wh{x}, k) + \delta\norm{2}{\wh{x}}
  \]
  in $O(\frac{k}{\eps} \log (n/k) \log (n/\delta))$ time.
\end{theorem}
\begin{proof}
  Define $f_r = O(1/r^2)$ so $\sum f_r < 1/4$.  Choose $R$ so
  $\prod_{r \leq R} f_r < 1/k \leq \prod_{r < R} f_r$.  Then $R =
  O(\log k / \log \log k)$, since $\prod_{r \leq R} f_r <
  (f_{R/2})^{R/2} = (2/R)^{R}$.

  Set $\eps_r = f_r \eps$, $\alpha_r = \Theta(f_r^2)$, $k_r =
  k\prod_{i < r} f_i$, $B_r = O(\frac{k}{\eps}\alpha_rf_r)$.  Then $B_r =
  \omega(\frac{k_r}{\alpha_r^2 \eps_r})$, so for sufficiently large
  constant the constraint of Lemma~\ref{lemma:sparsefftoneloop} is
  satisfied.  For appropriate constants,
  Lemma~\ref{lemma:sparsefftoneloop} says that in each round $r$,
  \begin{align}\label{e:errlink}
    \err^2(\wh{x}^{(r+1)}, k_{r+1}) &= \err^2(\wh{x}^{(r+1)}, f_r k_r)
    \leq (1+f_r\eps)\err^2(\wh{x}^{(r)}, k_r) +
    O(f_r\eps\delta^2n\norm{1}{\wh{x}}^2)
  \end{align}
  with probability at least $1 - f_r$.  The error accumulates, so in
  round $r$ we have
  \begin{align*}
    \err^2(\wh{x}^{(r)}, k_r) &\leq \err^2(\wh{x}, k)
    \prod_{i < r} (1 + f_i\eps) + 
    \sum_{i < r} O(f_r\eps\delta^2n\norm{1}{\wh{x}}^2) \prod_{i < j < r} (1 + f_j
    \eps)
  \end{align*}
  with probability at least $1 - \sum_{i < r} f_i > 3/4$.  Hence in the end,
  since $k_{R+1} = k\prod_{i \leq R} f_i < 1$,
  \begin{align*}
    \norm{2}{\wh{x}^{(R+1)}}^2 &= \err^2(\wh{x}^{(R+1)}, k_{R+1})
    \leq \err^2(\wh{x}, k) \prod_{i \leq R} (1 + f_i
    \eps) + O(R\eps\delta^2n\norm{1}{\wh{x}}^2) \prod_{i \leq R} (1 + f_i
    \eps)
  \end{align*}
  with probability at least $3/4$.  We also have
  \[
  \prod_i (1 + f_i \eps) \leq e^{\eps \sum_i f_i} \leq e
  \]
  making
  \[
  \prod_i (1 + f_i \eps) \leq 1 + e \sum_i f_i \eps < 1 + 2\eps.
  \]
  Thus we get the approximation factor
  \[
  \norm{2}{\wh{x}-\wh{z}^{(R+1)}}^2 \leq (1+2\eps) \err^2(\wh{x}, k) + O((\log k)\eps\delta^2 n \norm{1}{\wh{x}}^2)
  \]
  with at least $3/4$ probability.  Rescaling $\delta$ by
  $\text{poly}(n)$, using $\norm{1}{\wh{x}}^2 \leq n\norm{2}{\wh{x}}$,
  and taking the square root gives the desired
  \[
  \norm{2}{\wh{x}-\wh{z}^{(R+1)}} \leq (1+\eps) \err(\wh{x}, k) + \delta\norm{2}{\wh{x}}.
  \]
  Now we analyze the running time.  The update $\wh{z}^{(r+1)} -
  \wh{z}^{(r)}$ in round $r$ has support size $3k_r$, so in round $r$
  \[
  \snorm{0}{\wh{z}^{(r)}} \leq \sum_{i < r} 3k_r = O(k).
  \]
  Thus the expected running time in round $r$ is
  \begin{align*}
    &O((k(1 + \alpha_r \log (n/\delta)) +
    \frac{B_r}{\alpha_r} \log (n/\delta)) (\log \frac{1}{\alpha_r\eps_r} + \log (n/B_r)))\\
    ={}&O((k + \frac{k}{r^4}\log(n/\delta) + \frac{k}{\eps r^2} \log
    (n/\delta)) (\log\frac{r^2}{\eps} + \log (n\eps/k) + \log
    r))\\
    ={}& O((k + \frac{k}{\eps r^2} \log (n/\delta)) (\log r +
    \log (n/k)))
  \end{align*}
  We split the terms multiplying $k$ and $\frac{k}{\eps r^2} \log
  (n/\delta)$, and sum over $r$.  First,
  \begin{align*}
    \sum_{r=1}^R (\log r + \log (n/k))
    &\leq R \log R + R\log (n/k)\\
    &\leq O(\log k + \log k \log (n/k))\\
    &= O(\log k \log (n/k)).
  \end{align*}
  Next,
  \begin{align*}
    &\sum_{r=1}^R \frac{1}{r^2}(\log r + \log (n/k)) = O(\log (n/k))
  \end{align*}
  Thus the total running time is
  \begin{align*}
    O(k \log k \log (n/k) + \frac{k}{\eps} \log (n/\delta) \log (n/k))
    = O(\frac{k}{\eps} \log (n/\delta) \log (n/k)).
  \end{align*}
\end{proof}

\section{Reducing the full $k$-dimensional DFT to the exact $k$-sparse case in $n$ dimensions}
\label{s:red}
In this section we show the following lemma. Assume that $k$ divides $n$.

\begin{lemma}
  Suppose that there is an algorithm $A$ that, given an $n$-dimensional
  vector $y$ such that $\hat{y}$ is $k$-sparse, computes $\hat{y}$ in
  time $T(k)$. Then there is an algorithm $A'$ that given a
  $k$-dimensional vector $x$ computes $\hat{x}$ in time $O(T(k)))$.
\end{lemma}

\begin{proof} 
Given a $k$-dimensional vector $x$, we define $y_i = x_{i \bmod k}$, for $i = 0 \ldots n-1$. 
Whenever $A$ requests a sample $y_i$, we compute it from $x$ in constant time. 
Moreover, we have that $\hat{y}_i = \hat{x}_{i/(n/k)}$ if $i$ is a multiple of $(n/k)$, and $\hat{y}_i = 0$ otherwise. 
Thus $\hat{y}$ is $k$-sparse. Since $\hat{x}$ can be immediately recovered from $\hat{y}$, the lemma follows.
\end{proof}

\begin{corollary}
Assume that the $n$-dimensional DFT cannot be computed in $o(n \log n)$ time. 
Then any algorithm for the $k$-sparse DFT (for vectors of arbitrary dimension) must run in $\Omega(k \log k)$ time. 
\end{corollary}
\section{Lower Bound}\label{sec:lower}

In this section, we show any algorithm satisfying Equation~\eqref{e:l2l2} must
access $\Omega(k \log (n/k) / \log \log n)$ samples of $x$.

We translate this problem into the language of compressive sensing:

\begin{theorem}\label{thm:lowerbound}
  Let $F \in \C^{n \times n}$ be orthonormal and satisfy
  $\abs{F_{i,j}} = 1/\sqrt{n}$ for all $i, j$.  Suppose an algorithm
  takes $m$ adaptive samples of $Fx$ and computes $x'$ with
  \[
  \norm{2}{x-x'} \leq 2\min_{k\text{-sparse } x^*}\norm{2}{x-x^*}
  \]
  for any $x$, with probability at least $3/4$.  Then it must have $m
  = \Omega(k\log (n/k) / \log \log n)$.
\end{theorem}
\begin{corollary}
  Any algorithm computing the approximate Fourier transform must
  access $\Omega(k \log (n/k) / \log \log n)$ samples from the time
  domain.
\end{corollary}

If the samples were chosen non-adaptively, we would immediately have
$m = \Omega(k\log (n/k))$ by~\cite{PW}.  However, an algorithm could
choose samples based on the values of previous samples.  In the sparse
recovery framework allowing general linear measurements, this
adaptivity can decrease the number of measurements to $O(k \log \log
(n/k))$~\cite{IPW}; in this section, we show that adaptivity is much
less effective in our setting where adaptivity only allows the choice
of Fourier coefficients.

We follow the framework of Section~4 of~\cite{PW}.  In this section we
use standard notation from information theory, including $I(x; y)$ for
mutual information, $H(x)$ for discrete entropy, and $h(x)$ for
continuous entropy.  Consult a reference such as~\cite{CT}
for details.

Let $\mathcal{F} \subset \{S \subset [n] : \abs{S} = k\}$ be a
family of $k$-sparse supports such that:
\begin{itemize}
\item $\abs{S \oplus S'} \geq k$ for $S \neq S' \in \mathcal{F}$, where $\oplus$ denotes the exclusive difference between two sets, and
\item $\log \abs{\mathcal{F}} = \Omega(k \log (n/k))$.
\end{itemize}
This is possible; for example, a random code on $[n/k]^k$ with
relative distance $1/2$ has these properties.

For each $S \in \mathcal{F}$, let $X^S = \{x \in \{0, \pm 1\}^n \mid
\supp(x^S) = S\}$.  Let $x \in X^S$ uniformly at random.  The variables $x_i$, $i \in
S$,  are i.i.d. subgaussian random variables with
parameter $\sigma^2 = 1$, so for any row $F_j$ of $F$, $F_jx$ is
subgaussian with parameter $\sigma^2 = k/n$. Therefore
\[
\Pr_{x \in X^S}[\abs{F_jx} > t\sqrt{k/n}] < 2e^{-t^2/2}
\]
hence for each $S$, we can choose an $x^S \in X^S$ with
\begin{align}\label{e:lowerlinf}
  \norm{\infty}{Fx^S} < O(\sqrt{\frac{k\log n}{n}}).
\end{align}
Let $X = \{x^S \mid S \in \mathcal{F}\}$ be the set of such $x^S$.

Let $w \sim N(0, \alpha\frac{k}{n}I_n)$ be i.i.d. normal with variance
$\alpha k/n$ in each coordinate.

Consider the following process:

\paragraph{Procedure.} First, Alice chooses $S \in \mathcal{F}$
uniformly at random, then selects the $x \in X$ with $\supp(x) = S$.
Alice independently chooses $w\sim N(0, \alpha\frac{k}{n}I_n)$ for a
parameter $\alpha = \Theta(1)$ sufficiently small.  For $j\in [m]$,
Bob chooses $i_j \in [n]$ and observes $y_j = F_{i_j}(x + w)$.  He
then computes the result $x' \approx x$ of sparse recovery, rounds to
$X$ by $\hat{x} = \argmin_{x^* \in X} \norm{2}{x^* - x'}$, and sets
$S' = \supp(\hat{x})$.  This gives a Markov chain $S \to x \to y \to
x' \to \hat{x} \to S'$.

We will show that deterministic sparse recovery algorithms require
large $m$ to succeed on this input distribution $x+w$ with $3/4$
probability.  By Yao's minimax principle, this means randomized sparse recovery algorithms also
require large $m$ to succeed with $3/4$ probability.

Our strategy is to give upper and lower bounds on $I(S; S')$, the mutual information between $S$ and $S'$.

\begin{lemma}[Analog of Lemma 4.3 of~\cite{PW} for $\eps = O(1)$]
\label{lemma:inflower} There exists a constant $\alpha' > 0$
  such that if $\alpha < \alpha'$, then
  $I(S; S') = \Omega(k \log (n/k))$ .
\end{lemma}
\begin{proof}
  Assuming the sparse recovery succeeds (as happens with 3/4 probability), we have
$    \norm{2}{x'-(x+w)} \le 2 \norm{2}{w}$, which implies 
 $   \norm{2}{x'-x} \leq 3\norm{2}{w}$. Therefore
\begin{align*}
    \norm{2}{\hat{x}-x} &\leq \norm{2}{\hat{x}-x'} + \norm{2}{x'-x}\\
    &\leq 2\norm{2}{x'-x}\\
    &\leq 6\norm{2}{w}.
  \end{align*}
  We also know $\norm{2}{x'-x''} \geq \sqrt{k}$ for all distinct
  $x',x'' \in X$ by construction.  Because $\E[\norm{2}{w}^2] = \alpha
  k$, with probability at least $3/4$ we have $\norm{2}{w} \leq
  \sqrt{4\alpha k} < \sqrt{k}/6$ for sufficiently small $\alpha$.  But
  then $\norm{2}{\hat{x}-x} < \sqrt{k}$, so $\hat{x} = x$ and $S =
  S'$.  Thus $\Pr[S \neq S'] \leq 1/2$.

  Fano's inequality states $H(S \mid S') \leq 1 + \Pr[S \neq S'] \log
  \abs{\mathcal{F}}$.  Thus
  \[
  I(S; S') = H(S) - H(S \mid S') \geq -1 + \frac{1}{2} \log \abs{\mathcal{F}} = \Omega(k \log (n/k)) 
  \]
  as desired.
\end{proof}

We next show an analog of their upper bound (Lemma~4.1 of~\cite{PW})
on $I(S; S')$ for adaptive measurements of bounded $\ell_\infty$ norm.
The proof follows the lines of~\cite{PW}, but is more careful about
dependencies and needs the $\ell_\infty$ bound on $Fx$.

\begin{lemma}\label{lemma:infupper}
  \[
  I(S; S') \leq O(m \log (1 + \frac{1}{\alpha}\log n)).
  \]
\end{lemma}
\begin{proof}
  Let $A_j = F_{i_j}$ for $j \in [m]$, and let $w'_j = A_jw$.  The
  $w'_j$ are independent normal variables with variance $\alpha
  \frac{k}{n}$.  Because the $A_j$ are orthonormal and $w$ is drawn
  from a rotationally invariant distribution, the $w'$ are also
  independent of $x$.

  Let $y_j = A_jx+w'_j$.  We know $I(S; S') \leq I(x; y)$ because $S
  \to x \to y \to S'$ is a Markov chain.  Because the variables $A_j$
  are deterministic given $y_1, \dotsc, y_{j-1}$,
  \begin{align*}
    I(x; y_j \mid y_1, \dotsc, y_{j-1})
    &= I(x; A_jx + w'_j \mid y_1, \dotsc, y_{j-1})\\
    &= h(A_jx + w'_j \mid y_1, \dotsc, y_{j-1}) - h(A_jx + w'_j \mid x, y_1, \dotsc, y_{j-1})\\
    &= h(A_jx + w'_j \mid y_1, \dotsc, y_{j-1}) - h(w'_j).
  \end{align*}
  By the chain rule for information,
  \begin{align*}
    I(S; S') &\leq I(x; y)\\
    &= \sum_{j=1}^m I(x; y_j \mid y_1, \dotsc, y_{j-1})\\
    &= \sum_{j=1}^m h(A_jx + w'_j \mid y_1, \dotsc, y_{j-1}) - h(w'_j)\\
    &\leq \sum_{j=1}^m h(A_jx + w'_j) - h(w'_j).
  \end{align*}
  Thus it suffices to show $h(A_jx + w'_j) - h(w'_j) = O(\log (1 +
  \frac{1}{\alpha}\log n))$ for all $j$.

  Note that $A_j$ depends only on $y_1, \dotsc, y_{j-1}$, so it is
  independent of $w'_j$.  Thus
  \begin{align*}
    \E[(A_j x + w'_j)^2] = \E[(A_j x)^2] + \E[(w'_j)^2] \leq
    O(\frac{k\log n}{n}) + \alpha \frac{k}{n}
  \end{align*}
  by Equation~\eqref{e:lowerlinf}.  Because the maximum entropy
  distribution under an $\ell_2$ constraint is a Gaussian, we have
  \begin{align*}
    h(A_jx + w'_j) - h(w'_j)
    &\leq h(N(0, O(\frac{k\log n}{n}) + \alpha \frac{k}{n})) - h(N(0, \alpha \frac{k}{n}))\\
    &= \frac{1}{2}\log(1 + \frac{O(\log n)}{\alpha})\\
    &= O(\log(1 + \frac{1}{\alpha} \log n)).
  \end{align*}
  as desired.
\end{proof}

Theorem~\ref{thm:lowerbound} follows from Lemma~\ref{lemma:inflower}
and Lemma~\ref{lemma:infupper}, with $\alpha = \Theta(1)$.

\section{Efficient Constructions of Window Functions}\label{sec:efficientwindow}

\def\cdft{\widetilde{\cdf}}

\begin{claim}\label{claim:cdf}
  Let $\cdf$ denote the standard Gaussian cumulative distribution
  function.  Then:
  \begin{enumerate}
  \item $\cdf(t) = 1 - \cdf(-t)$.
  \item $\cdf(t) \leq e^{-t^2/2}$ for $t < 0$.
  \item $\cdf(t) < \delta$ for $t < -\sqrt{2\log (1/\delta)}$.
  \item $\int_{x=-\infty}^t \cdf(x)dx < \delta$ for $t < -\sqrt{2
      \log(3/\delta)}$.
  \item For any $\delta$, there exists a function
    $\cdft_{\delta}(t)$ computable in $O(\log(1/\delta))$ time such
    that $\norm{\infty}{\cdf - \cdft_{\delta}} < \delta$.
  \end{enumerate}
\end{claim}
\begin{proof}
  \ 
  \begin{enumerate}
  \item Follows from the symmetry of Gaussian distribution.
  \item Follows from a standard moment generating function bound on
    Gaussian random variables.
  \item Follows from (2).
  \item Property (2) implies that $\cdf(t)$ is at most $\sqrt{2\pi} <
    3$ times larger than the Gaussian pdf.  Then apply (3).
  \item By (1) and (3), $\cdf(t)$ can be computed as $\pm \delta$ or
    $1 \pm \delta$ unless $\abs{t} < \sqrt{2(\log(1/\delta))}$.  But
    then an efficient expansion around $0$ only requires
    $O(\log(1/\delta))$ terms to achieve precision $\pm \delta$.

    For example, we can truncate the representation~\cite{Mar04}
    \[
    \cdf(t) = \frac{1}{2} + \frac{e^{-t^2/2}}{\sqrt{2\pi}}\left(t  + \frac{t^3}{3} + \frac{t^5}{3 \cdot 5} + \frac{t^7}{3\cdot 5 \cdot 7} + \dotsb\right)
    \]
    at $O(\log (1/\delta))$ terms.
  \end{enumerate}
\end{proof}

\begin{claim}\label{claim:continuousdft}
  Define the continuous Fourier transform of $f(t)$ by
  \[
  \wh{f}(s) = \int_{-\infty}^\infty e^{-2\pi \mathbf{i} st}f(t)dt.
  \]
  For $t \in [n]$, define
  \[
  g_t = \sqrt{n}\sum_{j = -\infty}^\infty f(t + nj)
  \]
  and
  \[
  g'_t = \sum_{j = -\infty}^\infty \wh{f}(t/n + j).
  \]
  Then $\wh{g} = g'$, where $\wh{g}$ is the $n$-dimensional DFT of
  $g$.
\end{claim}
\begin{proof}
  Let $\Delta_1(t)$ denote the Dirac comb of period $1$: $\Delta_1(t)$
  is a Dirac delta function when $t$ is an integer and zero elsewhere.
  Then $\wh{\Delta_1} = \Delta_1$.  For any $t\in [n]$, we have
  \begin{align*}
    \wh{g}_t &= \sum_{s=1}^n \sum_{j=-\infty}^\infty f(s + nj) e^{-2\pi \mathbf{i} t s/n}\\
    &= \sum_{s=1}^n \sum_{j=-\infty}^\infty f(s + nj) e^{-2\pi \mathbf{i} t (s+nj)/n}\\
    &= \sum_{s=-\infty}^\infty f(s) e^{-2\pi \mathbf{i} t s/n}\\
    &= \int_{-\infty}^\infty f(s) \Delta_1(s) e^{-2\pi \mathbf{i} t s / n} ds \\
    &= \wh{(f \cdot \Delta_1)}(t/n)  \\
    &= (\wh{f} * \Delta_1)(t/n)\\
    &= \sum_{j=-\infty}^\infty \wh{f}(t/n + j)\\
    &= g'_t.
  \end{align*}

\end{proof}

\begin{lemma}
  For any parameters $B \geq 1, \delta > 0,$ and $\alpha > 0$, there
  exist flat window functions $G$ and $\wh{G'}$ such that $G$ can be
  computed in $O(\frac{B}{\alpha}\log(n/\delta))$ time, and for each
  $i$ $\wh{G'}_i$ can be evaluated in $O(\log (n/\delta))$ time.
\end{lemma}
\begin{proof}
  We will show this for a function $\wh{G'}$ that is a Gaussian
  convolved with a box-car filter.  First we construct analogous
  window functions for the continuous Fourier transform.  We then show
  that discretizing these functions gives the desired result.

  Let $D$ be the pdf of a Gaussian with standard deviation $\sigma > 1$ to be
  determined later, so $\wh{D}$ is the pdf of a Gaussian with standard deviation
  $1/\sigma$.  Let $\wh{F}$ be a box-car filter of length $2C$ for
  some parameter $C < 1$; that is, let $\wh{F}(t) = 1$ for $\abs{t} < C$
  and $F(t) = 0$ otherwise, so $F(t) = 2C\text{sinc}(t/(2C))$.  Let $G^* =
  D \cdot F$, so $\wh{G^*} = \wh{D} * \wh{F}$.

  Then $\abs{G^*(t)} \leq 2C\abs{D(t)} < 2C\delta$ for $\abs{t} >
  \sigma\sqrt{2\log (1/\delta)}$. Furthermore, $G^*$ is computable in
  $O(1)$ time.

  Its Fourier transform is $\wh{G^*}(t) = \cdf(\sigma(t + C)) -
  \cdf(\sigma(t - C))$.  By Claim~\ref{claim:cdf} we have for $\abs{t}
  > C + \sqrt{2\log (1/\delta)} / \sigma$ that $\wh{G^*}(t) = \pm
  \delta$.  We also have, for $\abs{t} < C -
  \sqrt{2\log(1/\delta)}/\sigma$, that $\wh{G^*}(t) = 1 \pm 2\delta$.

 Now, for $i \in [n]$ let $H_i = \sqrt{n}\sum_{j=\infty}^\infty
  G^*(i+nj)$. By Claim~\ref{claim:continuousdft} it has DFT $\wh{H}_i
  = \sum_{j=\infty}^\infty \wh{G^*}(i/n+j)$.  Furthermore,
  \begin{align*}
    \sum_{\abs{i} > \sigma \sqrt{2 \log (1/\delta)}} \abs{G^*(i)}
    &\leq 4C \sum_{i < -\sigma \sqrt{2 \log (1/\delta)}} \abs{D(i)}\\
    &\leq 4C \left(\int_{-\infty}^{-\sigma \sqrt{2 \log (1/\delta)}} \abs{D(x)}dx + D(-\sigma \sqrt{2 \log (1/\delta)})\right)\\
    &\leq 4C(\cdf(-\sqrt{2 \log (1/\delta)}) + D(-\sigma \sqrt{2 \log (1/\delta)}))\\
    &\leq 8C\delta \leq 8\delta.
  \end{align*}
  Thus if we let
  \[
  G_i = \sqrt{n}\sum_{\substack{\abs{j} < \sigma\sqrt{2\log (1/\delta)}\\j
      \equiv i \pmod n}}G^*(j)
  \]
  for $\abs{i} < \sigma\sqrt{2\log (1/\delta)}$ and $G_i = 0$
  otherwise, then $\norm{1}{G-H} \leq 8\delta\sqrt{n}$.

  Now, note that for integer $i$ with $\abs{i} \leq n/2$,
  \begin{align*}
    \wh{H}_i - \wh{G^*}(i/n) &= \sum_{\substack{j\in \Z\\j \neq 0}} \wh{G^*}(i/n + j)\\
    \abs{\wh{H}_i - \wh{G^*}(i/n)}&\leq 2 \sum_{j = 0}^{\infty} \wh{G^*}(-1/2 - j)\\
    &\leq 2 \sum_{j = 0}^{\infty} \cdf(\sigma(-1/2 - j + C))\\
    &\leq 2 \int_{-\infty}^{-1/2} \cdf(\sigma(x + C)) dx +
    2\cdf(\sigma(-1/2 + C))\\
    &\leq 2 \delta / \sigma + 2 \delta \leq 4\delta
  \end{align*}
  by Claim~\ref{claim:cdf}, as long as
  \begin{align}\label{e:requirement}
    \sigma(1/2 - C) > \sqrt{2 \log (3/\delta)}.
  \end{align}
  Let
  \[
  \wh{G'}_i = \left\{
    \begin{array}{cl}
      1 & \abs{i} \leq n(C - \sqrt{2 \log (1/\delta)}/\sigma)\\
      0 & \abs{i} \geq n(C + \sqrt{2 \log (1/\delta)}/\sigma)\\
      \cdft_\delta(\sigma(i+C)/n) - \cdft_\delta(\sigma(i-C)/n) & \text{otherwise}
    \end{array}
  \right.
  \]
  where $\cdft_\delta(t)$ computes $\cdf(t)$ to precision $\pm \delta$
  in $O(\log (1/\delta))$ time, as per Claim~\ref{claim:cdf}.  Then
  $\wh{G'}_i = \wh{G^*}(i/n) \pm 2\delta = \wh{H}_i \pm 6\delta$.
  Hence
  \begin{align*}
    \norm{\infty}{\wh{G}-\wh{G'}} &\leq \norm{\infty}{\wh{G'}-\wh{H}} +
    \norm{\infty}{\wh{G}-\wh{H}} 
    \\&
    \leq \norm{\infty}{\wh{G'}-\wh{H}} +
    \norm{2}{\wh{G}-\wh{H}} 
    \\&
    = \norm{\infty}{\wh{G'}-\wh{H}} + \norm{2}{G-H}
    \\&
    \leq \norm{\infty}{\wh{G'}-\wh{H}} + \norm{1}{G-H} \\
    &\leq (8\sqrt{n} + 6) \delta.
  \end{align*}
  Replacing $\delta$ by $\delta/n$ and plugging in $\sigma =
  \frac{4B}{\alpha} \sqrt{2\log (n/\delta)} > 1$ and $C =
  (1-\alpha/2)/(2B) < 1$, we have the required properties of flat
  window functions:
  \begin{itemize}
  \item $\abs{G_i} = 0$ for $\abs{i} \geq
    \Omega(\frac{B}{\alpha}\log(n/\delta))$
  \item $\wh{G'}_i = 1$ for $\abs{i} \leq (1-\alpha)n/(2B)$
  \item $\wh{G'}_i = 0$ for $\abs{i} \geq n/(2B)$
  \item $\wh{G'}_i \in [0, 1]$ for all $i$.
  \item $\norm{\infty}{\wh{G'}-\wh{G}} < \delta$.
  \item We can compute $G$ over its entire support in
    $O(\frac{B}{\alpha}\log(n/\delta))$ total time.
  \item For any $i$, $\wh{G'}_i$ can be computed in
    $O(\log(n/\delta))$ time for $\abs{i} \in [(1-\alpha)n/(2B),
    n/(2B)]$ and $O(1)$ time otherwise.
  \end{itemize}
  The only requirement was Equation~\eqref{e:requirement}, which is that
  \[
  \frac{4B}{\alpha}\sqrt{2 \log (n/\delta)}(1/2 -
  \frac{1-\alpha/2}{2B}) > \sqrt{2 \log (3n/\delta)}.
  \]
  This holds if $B \geq 2$.  The $B = 1$ case is trivial using the
  constant function $\wh{G'}_i = 1$.
\end{proof}

\section{Open questions}\label{sec:open}

\begin{itemize}
\item  Design an $O(k \log n)$-time algorithm for general signals. 
Alternatively, prove that no such algorithm exists, under
 ``reasonable'' assumptions.\footnote{The $\Omega(k \log
 (n/k)/\log \log n)$ lower bound for the sample complexity shows that
 the running time of our algorithm, $O(k \log n \log(n/k))$, is equal
 to the sample complexity of the problem times (roughly) $\log n$. One
 could speculate that this logarithmic discrepancy is due to the need
 for using FFT to process the samples. Although we do not have any
 evidence for the optimality of our general algorithm, the ``sample
 complexity times $\log n$'' bound appears to be a natural barrier to
 further improvements.} 
\item Reduce the sample complexity of the
 algorithms. Currently, the number of samples used by each algorithm
 is only bounded by their running times.
\item Extend the results to other (related) tasks, such as computing 
the sparse Walsh-Hadamard Transform.
\item Extend the algorithm to the case when $n$ is not a power of
  $2$. Note that some of the earlier algorithms, e.g.,~\cite{GMS},
  work for any $n$.
\item Improve the failure probability of the algorithms.  Currently,
  the algorithms only succeed with constant probability.
  Straightforward amplification would take a $\log(1/p)$ factor
  slowdown to succeed with $1-p$ probability. One would hope to avoid
  this slowdown.
 \end{itemize} 

\section*{Acknowledgements}

The authors would like to thank Martin Strauss  and Ludwig Schmidt for many helpful
comments about the writing of the paper. This work is supported by the
Space and Naval Warfare Systems Center Pacific under Contract
No. N66001-11-C-4092, David and Lucille Packard Fellowship, and NSF
grants CCF-1012042 and CNS-0831664. E. Price is supported in part by
an NSF Graduate Research Fellowship.

\bibliographystyle{alpha}
\bibliography{paper}

\end{document}